# Directed Energy Missions for Planetary Defense

Philip Lubin[a], Gary B. Hughes[b], Mike Eskenazi[c], Kelly Kosmo[a], Isabella E. Johansson[a], Janelle Griswold[a], Mark Pryor[d], Hugh O'Neill[e], Peter Meinhold[a], Jonathon Suen[a], Jordan Riley[a], Qicheng Zhang[a], Kevin Walsh[f], Carl Melis[g], Miikka Kangas[a], Caio Motta[a], and Travis Brashears[a]

Corresponding author for questions:
lubin@deepspace.ucsb.edu

[a]Physics Department, University of California, Santa Barbara, CA 93106, USA
[b]Statistics Department, California Polytechnic State University, San Luis Obispo, CA 93407, USA
[c]ATK Space, 600 Pine Avenue, Goleta, CA 93117, USA
[d]Vorticy Inc, San Diego, CA 92121, USA
[e]Physics Department, Ventura College, Ventura, CA 93003, USA
[f]Southwest Research Institute, Boulder, CO 80302, USA
[g]Center for Astrophysics and Space Sciences, UC San Diego, San Diego, CA 92093, USA

For more information and related articles, videos and talks see:
http://www.deepspace.ucsb.edu/projects/directed-energy-planetary-defense

## ABSTRACT

Directed energy for planetary defense is now a viable option and is superior in many ways to other proposed technologies, being able to defend the Earth against all known threats. This paper presents basic ideas behind a directed energy planetary defense system that utilizes laser ablation of an asteroid to impart a deflecting force on the target. A conceptual philosophy called DE-STAR, which stands for **Directed Energy System for Targeting of Asteroids and exploRation**, is an orbiting stand-off system, which has been described in other papers. This paper describes a smaller, stand-on system known as **DE-STARLITE** as a reduced-scale version of DE-STAR. Both share the same basic heritage of a directed energy array that heats the surface of the target to the point of high surface vapor pressure that causes significant mass ejection thus forming an ejection plume of material from the target that acts as a rocket to deflect the object. This is generally classified as laser ablation. DE-STARLITE uses conventional propellant for launch to LEO and then ion engines to propel the spacecraft from LEO to the near-Earth asteroid (NEA). During laser ablation, the asteroid itself provides the propellant source material; thus a very modest spacecraft can deflect an asteroid much larger than would be possible with a system of similar mission mass using ion beam deflection (IBD) or a gravity tractor. DE-STARLITE is capable of deflecting an Apophis-class (325 m diameter) asteroid with a 1- to 15-year targeting time (laser on time) depending on the system design. The mission fits within the rough mission parameters of the Asteroid Redirect Mission (ARM) program in terms of mass and size. DE-STARLITE also has much greater capability for planetary defense than current proposals and is readily scalable to match the threat. It can deflect all known threats with sufficient warning.

**Keywords:** Directed Energy, Laser Phased Array, Planetary Defense, DE-STAR, DE-STARLITE

# 1. INTRODUCTION

**1.1 DE-STAR and DE-STARLITE**

While implementing a realistic directed energy planetary defense system may have seemed preposterous as little as a decade ago, recent technological developments allow serious consideration of such a system. The critical items such as phase locked laser amplifiers and lightweight photovoltaic deployable arrays are becoming increasingly more efficient and lower in mass. The necessary technology now exists to build such a system that will considerably enhance our ability to augment or enhance other methods to fulfill the need for planetary defense against asteroids that pose a threat of impacting Earth.

This paper primarily focuses on a design for a stand-on directed energy planetary defense system called **DE-STARLITE**. DE-STARLITE is a stand-on system, *i.e.*, it is designed to be delivered to a position that is nearby a threatening asteroid with a modest spacecraft and then work slowly on the threat to change its orbit. DE-STARLITE is suitable for mitigating targets that are many hundreds of meters in diameter and whose orbit is known to be a threat long before projected impact.

DE-STARLITE is one component of a more far-reaching philosophy for directed energy planetary defense. A future orbiting system is envisioned for stand-off planetary defense. The conceptual system is called DE-STAR, for **Directed Energy System for Targeting of Asteroids and exploRation**. Fluctuations in the Earth's atmosphere significantly hinder ground-based directed energy systems; thus, deploying a directed energy system above Earth's atmosphere eliminates such disturbances, as the interplanetary medium is not substantial enough to significantly affect the coherent beam. DE-STAR is discussed extensively in other papers (Lubin and Hughes, 2015; Kosmo *et al.*, 2015; Lubin *et al.*, 2014). The broader DE-STAR system is not discussed in depth in this paper, which will focus on DE-STARLITE.

**1.2 General Concepts for Orbit Deflection**

Residents near Chelyabinsk, Russia experienced the detrimental effects of a collision with a near-Earth asteroid (NEA) on 15 February 2013 as a ~20 m object penetrated the atmosphere above that city (Popova *et al.*, 2013). The effective yield from this object was approximately 1/2 Megaton TNT equivalent (Mt), or that of a large strategic warhead. The 1908 Tunguska event, also over Russia, is estimated to have had a yield of approximately 15 Mt and had the potential to kill millions of people had it come down over a large city (Garshnek *et al.*, 2000). Asteroid impacts pose a clear threat and future advancement to minimize this threat requires effective mitigation strategies.

A wide array of concepts for asteroid deflection has been proposed. Several detailed surveys of threat mitigation strategies are available in the literature, including Sanchez-Quartielles *et al.* (2007), Belton *et al.* (2004), Gritzner and Kahle (2004), and Morrison *et al.* (2002). Currently proposed diversion strategies can be broadly generalized into six categories.

(1a) Kinetic impactors, without explosive charges. An expendable spacecraft would be sent to intercept the threatening object. Direct impact could break the asteroid apart (Melosh and Ryan, 1997), and/or modify the object's orbit through momentum transfer. The energy of the impact could be enhanced via retrograde approach, *e.g.* McInnes (2004).

(1b) Kinetic impactors, with explosive charges. Momentum transfer using an expendable spacecraft could also be enhanced using an explosive charge, such as a nuclear weapon, *e.g.* Koenig and Chyba (2007).

(2) Gradual orbit deflection by surface albedo alteration. The albedo of an object could be changed using paint, *e.g.* Hyland *et al.* (2010). As the albedo is altered, a change in the object's Yarkovsky thermal drag would gradually shift the object's orbit. Similar approaches seek to create an artificial Yarkovsky effect, *e.g.* Vasile and Maddock (2010).

(3) Ion beam deflection (IBD) or ion beam shepherd (IBS) where high speed ions, such as the type used for ion thrusters, are directed at the asteroid from a nearby spacecraft, to push on asteroid and thus deflect it. (Bombardelli *et al.*, 2016; Brophy, 2015; Bombardelli, *et al.*, 2013; Bombardelli and Peláez, 2011).

(4) Direct motive force, such as by mounting a thruster directly to the object. Thrusters could include chemical propellants, solar or nuclear powered electric drives, or ion engines (Walker *et al.*, 2005).

(5) Indirect orbit alteration, such as gravity tractors. A spacecraft with sufficient mass would be positioned near the object, and maintain a fixed station with respect to the object using onboard propulsion. Gravitational attraction would tug the object toward the spacecraft, and gradually modify the object's orbit (Mazanek, *et al.*, 2015; Wie, 2008; Wie, 2007; McInnes, 2007; Schweickart *et al.*, 2006; Lu and Love, 2005).

(6) Expulsion of surface material such as by robotic mining. A robot on the surface of an asteroid would repeatedly eject material from the asteroid. The reaction force when material is ejected affects the object's trajectory (Olds *et al*., 2007).

(7) Vaporization of surface material. Like robotic mining, vaporization on the surface of an object continually ejects the vaporized material, creating a reactionary force that pushes the object into a new path. Vaporization can be accomplished by solar concentrators (Vasile and Maddock, 2010), lasers deployed from the ground (Phipps, 2010), or lasers deployed on spacecraft stationed near the asteroid (Maddock *et al*., 2007; Park and Mazenek, 2005; Gibbings *et al*., 2013; Phipps and Michaelis, 1994; Campbell, 2000; Vasile *et al*., 2013). One study (Kahle *et al*., 2006) envisioned a single large reflector mounted on a spacecraft traveling alongside an asteroid. The idea was expanded to a formation of spacecraft orbiting in the vicinity of the asteroid, each equipped with a smaller concentrator assembly capable of focusing solar power onto an asteroid at distances near ~1 km (Vasile and Maddock, 2010).

## 1.3 DE-STARLITE

The DE-STARLITE mission design, which is detailed in this paper, utilizes the same technologies and laser system as the larger standoff directed energy system. Namely, DE-STAR is a modular phased array of lasers that heat the surface of potentially hazardous asteroids to approximately 3000 K, a temperature sufficient to vaporize all known constituent materials. Mass ejection due to vaporization causes a reactionary force large enough to alter the asteroid's orbital trajectory and thus mitigate the risk of impact. Each DE-STAR system is characterized by the log of its linear size (Lubin *et al*., 2014). DE-STARLITE is basically a DE-STAR 0, consisting of a laser phased array on the order of 1 m in diameter. DE-STARLITE utilizes deployable photovoltaic arrays to power the system. A conceptual design is illustrated in Fig. 1.

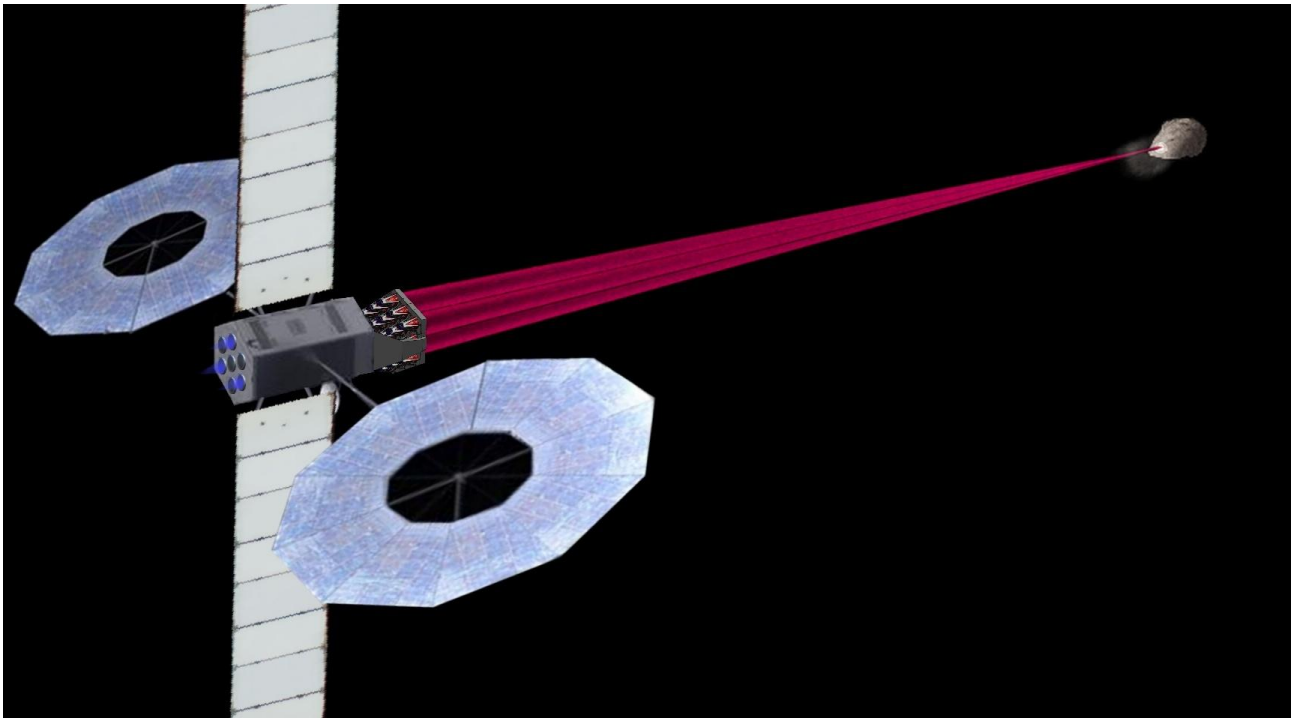

**Figure 1.** Artistic rendering of a deployed DE-STARLITE spacecraft deflecting an asteroid. The spacecraft is outfitted with two 15 m diameter MegaFlex PV Arrays, a z-folded radiator deployed up and down, a laser array mounted on a gimbal at the front, and ion engines at the back. The laser array can be either a phased array or a parallel non phased array. A baseline mission includes 1 m total aperture, with a goal to produce a 10 cm spot from a distance of 10 km. This is an artistic rendering only, demonstrating the overall concept. Note that the optimal thrust vector for orbit deflection is generally parallel or anti-parallel. One advantage of this approach is the ability to target the asteroid from a significant distance, mitigating the effects of the ejecta on the spacecraft. In addition to the ion engines shown in the rear of the spacecraft, there are small ion engines on the sides of the spacecraft for station keeping and maneuvering toward or away from the asteroid.

DE-STARLITE fits into the same basic launch vehicle and mass envelope as the current Asteroid Redirect Mission (ARM) block 1 program, which is designed to capture a 5-10 m diameter asteroid; however, DE-STARLITE is designed to be a true planetary defense system capable of redirecting large asteroids. It has been designed to use the same ion engines as the ARM program and the same PV system, though due to the reduced mass of DE-STARLITE, a much larger PV array can be deployed within the SLS block 1 mass allocation (70 tons to LEO) if desired. The scaling to megawatt class systems is discussed below. This paper will focus on a 100 kW (electrical) baseline DE-STARLITE as a feasible and fundable option that could pave the way for the ultimate long-term goal of a full standoff planetary defense system. Larger systems are also discussed. This paper details the design of the main elements of the spacecraft, namely, the photovoltaic panels, ion engines, laser array, and radiator as well as the parameters of the launch vehicles under consideration, and details the deflection capabilities of the system.

## 2. DESIGN

The objective is to design a system that will enable a spacecraft with a 1 m to 4.5 m diameter laser phased array to arrive at an NEA (Near Earth Asteroid) and deflect it from its potentially hazardous trajectory. The laser phased array is detailed in section 2.3, along with a lower risk potential fallback—a close packed focal plane array of fiber lasers. The propulsion for the LEO to NEA portion of the DE-STARLITE mission is made possible with a high-power solar electric propulsion (SEP) system (Brophy and Muirhead, 2011). The solar PV arrays, detailed in Section 2.1, convert power from the sun to provide system power. PV panels will originally be stowed for launch and will deploy upon reaching low-Earth orbit to provide a required 100 kW electrical power from two 15 m diameter ATK MegaFlex panels. Even larger power is possible within the launch mass and shroud sizes available as discussed in detailed below. The system will utilize ion engines (detailed in section 2.2) to propel the spacecraft from LEO to an NEA, as proposed in JPL's ARM program. The system aims to stay within the same mass and launch constraints as ARM and use much of the same propulsion technology. The laser efficiency determines the laser power obtained from the PV arrays; 35 kW of laser power would be produced at 35% efficiency and 50 kW at 50%.. The 35 kW estimate is based on the current efficiency (35%) of existing technology of the baseline Ytterbium laser amplifiers and thus provides for the worst case, while the 50 kW estimate is based on near term technological improvement within the next 5 years. A passive cooling radiator with z-folded arrays will be used to reject waste heat and maintain the temperature at an operational 300 K.

The basic design principle is to utilize a cylindrical bus with the lateral center of gravity close to the centerline (Kosmo *et al*., 2014). PV panels will be stowed at the back of the bus until deployment, and the hexagonal laser array will be mounted on a gimbal at the front of the spacecraft (Fig. 2). Radiator panels will deploy up and down (perpendicular to the bus) and will rotate about their axis so as to remain perpendicular to the sun in order to maximize radiator efficiency. Ion engines are located at the back of the spacecraft. Critical components are outlined in sections 2.1 through 2.4.

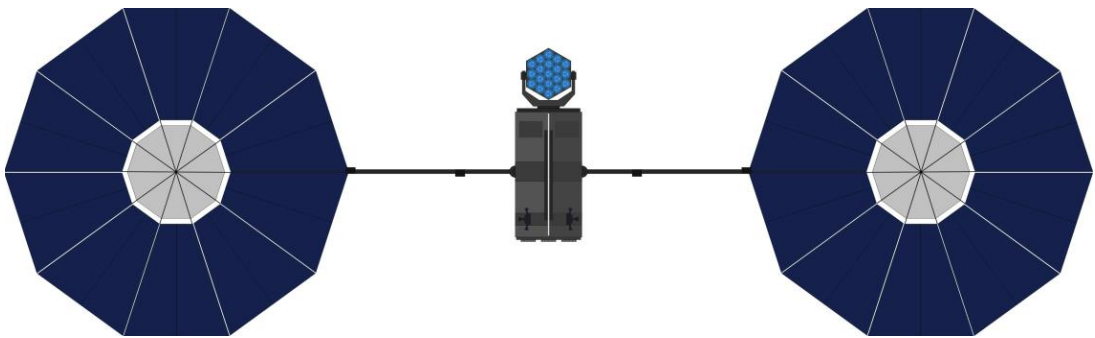

**Figure 2.** Conceptual design of the deployed spacecraft with two 15 m PV arrays that produce 50 kW each at the beginning of life for a total of 100 kW electrical, ion engines at the back, and the laser array pointed directly at the viewer. A 2 m diameter laser phased array is shown with 19 elements, each of which is 1-3 kW optical output. A 2 m diameter optical system is one of the possibilities for DE-STARLITE. More elements are easily added to allow for scaling to larger power levels. A 1 - 4.5 m diameter is feasible; no additional deflection comes from the larger optic, just additional range from the target.

## 2.1 Photovoltaic Panels

Two 15 m diameter MegaFlex PV arrays, manufactured by ATK Aerospace Systems in Goleta, CA, will be used to obtain the baselined 100 kW power solution. Extensive testing has been conducted on MegaFlex technology and the MegaFlex arrays have a high TRL (Murphy *et al*., 2014). Fig. 3 shows the PV array design and implementation.

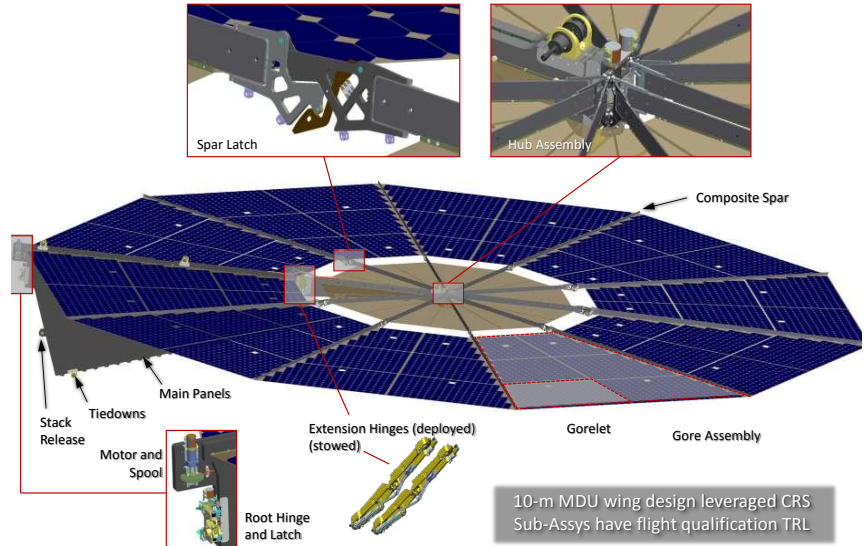

**Figure 3.** Detailed design of deployed MegaFlex array. Image courtesy ATK (Murphy *et al*., 2014).

## 2.2 Ion Engines

The spacecraft carrying DE-STARLITE will utilize ion propulsion to spiral out from LEO to its target (Kosmo *et al*., 2015). Ion engines are proposed for DE-STARLITE because they are between five and ten times more efficient than engines using conventional chemical propellants, depending on the type of ion engine.

## 2.3 Laser Array

The objective of the laser directed energy system is to project a large enough flux onto the surface of a near-Earth asteroid (via a highly focused coherent beam) to heat the surface to a temperature that exceeds the vaporization point of constituent materials, namely rock, as depicted in Fig. 4. This requires temperatures that depend on the material, but are typically around 2000-3000 K, or a flux in excess of $10^7$ W/m$^2$. A reactionary thrust due to mass ejection will divert the asteroid's trajectory (Lubin *et al*., 2014). To produce a great enough flux, the system must have both adequate beam convergence and sufficient power. From a distance of 10 km, a spot size on the asteroid of 10 cm provides enough flux to vaporize (sublimate) rock (Hughes *et al*., 2014). Optical aperture size, pointing control and jitter, and efficacy of adaptive optics techniques are several critical factors that affect beam convergence. As mentioned, the optical power output of the laser is projected to be between 35 kW and 70 kW, depending on technological advancements in laser amplifier efficiency in the coming years. Currently the amplifiers are about 35% efficient but it is expected they will exceed 50% within five years. Similar requirements are sought by power beaming systems (Mankins, 1997; Lin 2002). For the optional (non-phase-locked) fiber focal plane array the lasers are even more efficient and already exceed 50%. Any power level in this range will work for the purpose of this mission, but higher efficiency allows for more thrust on the target for a given electrical input as well as for smaller radiators and hence lower mission mass.

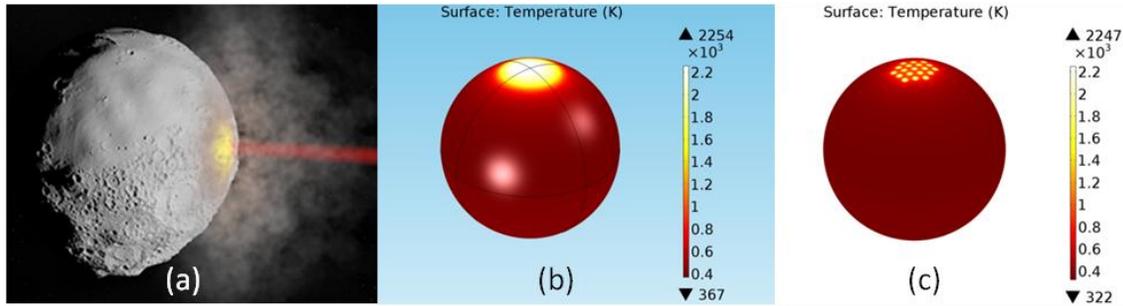

**Figure 4**. **LEFT**: Physics based simulation of asteroid laser ablation (Lubin *et al*., 2014). **MIDDLE**: Simulation showing one spot from the baseline phased array on the target at sufficient temperature to cause ablation. **RIGHT**: Multi-beam simulation depicting 19 beams on the target from an optional choice of a close packed fiber laser focal plane array.

The proposed baseline optical system consists of 19 individual optical elements in a phased array. A single element concept is shown in Fig. 5. A significant benefit of utilizing an array of phase-locked laser amplifiers is that it is completely modular and thus scalable to much larger systems, and allows for a greater range than would a close packed array with a single optic. Focusing and beam steering are achieved by controlling the relative phase of individual laser elements. Rough pointing of the array to the target is determined by spacecraft attitude control and gimbal pointing of the optics. Laser tips behind each optical element are mounted on 6-axis micro-positioner hexapod; lateral movement of the laser tips behind each lens provides intermediate pointing adjustment for individual array elements. Each fiber tip is supported on the hexapod and can be augmented with a z-axis rapid position controller if needed. It is not clear if this is needed currently. Precision beam steering is accomplished by coordinated phase modulation across the array by z-position control of the fiber tips as well as by electronic phase modulation. Each fiber is fed with a phase-controllable laser amplifier. Phase feedback from in front of the lens array to each phase controller provides a signal for beam formation adjustment (spot focus). Phase alignment is maintained to within lambda/10 1-sigma RMS across the entire array, assuming adequate phase controller system response (Hughes *et al*., 2014).

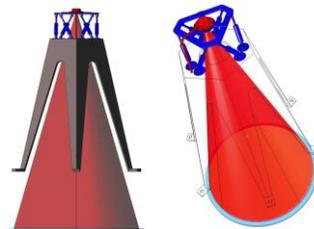

**Figure 5.** Single element of laser phased array, showing fiber-tip actuator for mid-level pointing control and rough phase alignment.

The DE-STARLITE spacecraft will be capable of outfitting a laser array with an aperture between 1 m and 4.5 m. Changing the aperture would not change the power of the laser; thus the solar arrays and radiator could be of the same dimensions, regardless. The difference in mass between the 1 m array and the 4.5 m array is not significant enough to pose new constraints. The benefit of a larger aperture is that the range of the laser from the target for a given power scales with the linear size of the optical aperture. For example, the range of a 3 m array is three times greater than that of a 1 m array. The benefit of having a longer range is that it allows the spacecraft to remain clear of the debris of the ejected material. The debris flux (kg/m$^2$ s) that hits the spacecraft drops as the square of the distance to the target. However, the total amount of particle debris (kg/s) on the optic is independent of distance since the range is proportional to the optic diameter, and the area of the optic is proportional to the square of the diameter. The main drawback of the larger aperture is a higher associated cost. The decision, thus, is dependent on funding and other mission specifics. Even sub-meter diameter optics are feasible if needed for specific missions.

The laser array will be placed on a gimbal to eliminate any potential issues with fuel usage in maneuvering the spacecraft, as depicted in Fig. 6. Further, it will allow for much greater flexibility in mission execution. This is imperative because the laser will have to raster scan the asteroid in order to maximize thrust, prevent burn through, and de-spin the asteroid if needed. Though much of this can be done with electronic steering, using a gimbal will be more energetically efficient than pointing the spacecraft. A gimbal would also be beneficial in the event that the spacecraft needs to orbit the target. Further, the added flexibility due to the gimbal mitigates risk by allowing the system to target smaller pieces of the asteroid that may get dislodged and pose a threat to the spacecraft. The gimbal will allow for two degrees of freedom because the angular orientation around the boresight of the spot on the asteroid is not a significant concern. This will be cheaper, easier to manufacture, and lighter than a system with greater degrees of freedom.

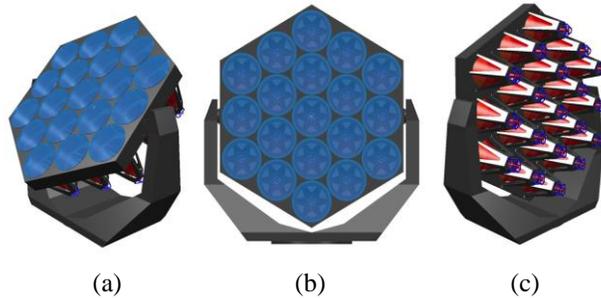

(a)        (b)        (c)

**Figure 6.** Mounted hexagonal laser phased array with a baseline of 19 elements depicted: **(a)** at 45 degrees, **(b)** face on, and **(c)** from the back.

### 2.4 Secondary Optical Arrangement

If necessary, a fallback option is to implement a hexagonal close packed focal plane array of laser fibers with a conventional optic such as a reflecting telescope instead of a phased array. A conceptual diagram is shown in Fig. 7. This system would consist of 19 circular fibers, each 25 μm in diameter with a sheath (cladding) around the inner core. The cladding will be 37.5 μm thick so that the center-to-center spacing of adjacent laser fibers is 100 μm. The thickness of the cladding may be increased if power leakage and cross talk is an issue. As with the phased array design, each fiber is attached to an amplifier; however, the fibers are close packed in the focal plane and utilize one larger hexapod and the lasers are NOT phase locked for simplicity. The close packed array will produce 19 individual spots on the target, separated center to center by the ratio of the target distance to optical size times the fiber spacing in the focal plane. For the baseline of 1.5 kW per amplifier, each fiber will illuminate the target with a spot diameter of approximately 12 mm and a center to center spacing of approximately 50 mm; however, this can be changed depending on the optical design. This option carries a lower risk, higher initial TRL, lower cost, and can also be implemented more rapidly. In addition, it requires the spacecraft to be significantly closer to the target than would be required with a phased array. The plan is to pursue both the phased array and the close packed array, and down select depending on specific mission parameters.

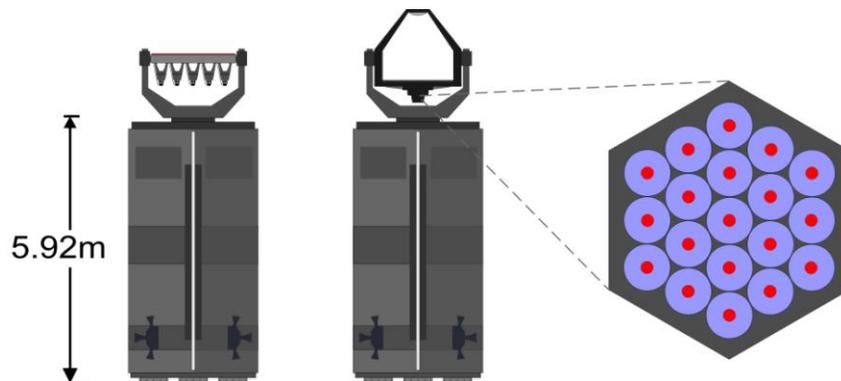

**Figure 7**. **LEFT**: Comparison view of mounted laser phased array and close packed array. **RIGHT**: Hexagonal close packed focal plane array of 19 laser fibers in the focal plane. Laser fibers have a diameter of 25 μm, the cladding around each of which is 37.5 μm thick.

### 2.5 Radiator

Thermal radiators are essential to spacecraft design so as to minimize incident radiation and maintain the spacecraft and its components at a functional temperature. The efficiency of the radiator can be determined by equation 1:

$$F = \dot{Q}/A = \varepsilon \sigma T^4 \qquad (1)$$

where $\varepsilon$ is the emittance of the surface, $\sigma$ is the Stephan-Boltzmann constant, $T$ is the temperature, $\dot{Q}$ is the heat rejected, $A$ is the area, and $F$ is the flux (Aaron, 2002). The baseline radiator will be coated in AZ-93 white paint, which has a high emittance of 0.91± 0.02 (or conservatively, 0.89) and a low alpha, as it only absorbs 14 -16% of incident sunlight on the spacecraft. The goal is to maintain a temperature of 300 K, as both the laser and onboard control electronics are operational at this temperature. At this temperature, the radiator can reject an idealized outward flux of 408 W/m$^2$. When taking into account the incident radiation, using a solar constant of 1362 W/m$^2$ and a maximum 16% absorptance, the net flux of energy across the surface of the radiator is approximately 190 W/m$^2$. The baseline is to prevent direct solar illumination of the radiator.

The area of the radiator must be determined by thermal analysis, and is dependent on the desired operating temperature, heating from the environment, interactions with other surfaces of the spacecraft (*e.g.*, solar arrays), and the highest estimate (worst case) satellite waste heat. The waste heat in this case is dependent on the efficiency of the laser amplifiers—35% or 50% as mentioned. The worst-case estimate (35% efficiency) requires 65 kW to be rejected as waste heat for a 100 kW electrical input assuming virtually all the power goes to the laser (which is approximately correct during laser firing). The required area $A$ can be easily determined:

$$\dot{Q}_{\text{rejected}} = AF_{\text{net}} \qquad (2)$$

where $F_{\text{net}}$ is the net outward flux and $\dot{Q}$ is the heat rejected. Given these parameters, the maximum required area of the radiator is ~341 m$^2$ for a 35% efficient laser amplifier. For a 50% efficient laser, a radiator area of ~262 m$^2$ is required. We assume that either a pumped liquid cooling loop or an advanced heatpipe would be used to transfer the heat from the laser to the radiator as is currently done now in the other uses of these laser amplifiers.

A passive cooling z-folded radiator consisting of two deployable panels will be used in order to provide a sufficient surface area over which to emit the waste heat generated by the system. Each panel z-folds out into six segments, each of which further folds out into two additional segments, making 18 segments in total for each panel. The panels will rotate about their axes to maximize efficiency by remaining perpendicular to the sun and by radiating out of both sides. Each segment will be 2.2 m by 2.2 m, yielding a total area of 348 m$^2$. Note that the radiators radiate out of both sides and that there are two radiator panels. These values are approximate; a more detailed radiator design would be required as part of an overall mission design. We would expect that, by the time of any mission start, significant increases in laser efficiency will have been achieved, thus reducing the required radiator size. Sun shades may also be used to limit solar absorption and thus allow for greater efficiency. The current mass to power ratio for radiators is about 25 kg/kW for the ARM system.

## 3. LAUNCH SYSTEM

The objective is to assess which launch vehicle is the most feasible and will provide the greatest performance given the mission directives of DE-STARLITE. The launch systems in consideration are Atlas V 551, Space Launch System (SLS) Block 1, Falcon Heavy, or Delta IV Heavy. These are likewise the launch systems in consideration for JPL's Asteroid Redirect Mission, which calls for a payload of comparable parameters (Brophy and Muirhead, 2013).

**Table 1**. Parameters of various launch vehicles in consideration for DE-STARLITE

| Parameter | Atlas V 551 | SLS Block 1 | Falcon Heavy | Delta IV Heavy |
|---|---|---|---|---|
| Payload Mass to LEO | 18 500 kg | 70 000 kg | 53 000 kg | 28 790 kg |
| Cost per unit mass to send into LEO | $13 200 / kg | $18 700 / kg | $1 890 / kg | $13 000 / kg |
| Diameter of Payload Fairing | 5.4 m | 8.4 m | 5.2 m | 5 m |
| Status | Flight proven | Development—First Expected Flight: 2017 | Development—First Expected flight: 2015 | Flight proven |

The DE-STARLITE spacecraft will fit within the payload fairing of any of the proposed launch systems, as depicted in Fig. 8. As is evident from the data in Table I, the SLS Block 1 has the highest capabilities, though also requires the highest cost. The Falcon Heavy demands the smallest cost per unit mass, and has capabilities between that of the Atlas V and SLS Block 1. While the Atlas V 551 and Delta IV Heavy have previously undergone successful missions, the SLS Block 1 and Falcon Heavy are projected to be flight-proven within the timescale of the DE-STARLITE mission. As with the Asteroid Redirect Mission, it is possible to compensate for the lower capabilities of the Atlas V by using the SEP system to spiral out of Earth's orbit and escape from Earth using Lunar Gravity Assist (LGA); however, this process of spiraling out and using LGA will take an additional 1 – 1.5 years of flight. All of these factors must be taken into consideration to choose the most effective launch system for the DE-STARLITE mission.

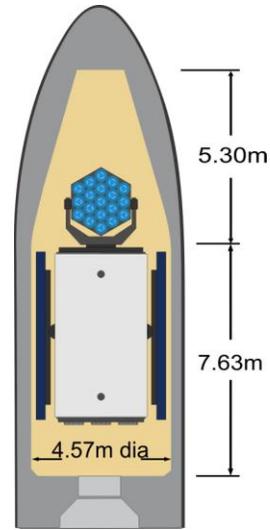

**Figure 8.** Stowed view of the DE-STARLITE spacecraft, in a 5 m fairing.

## 4. EXTENSIBILITY TO MEGAWATT SCALES

Both the laser array and the PV arrays are easily extended to larger power levels. The mass per unit power of the laser amplifiers is about 5 kg/kW currently with a strong push to bring this down to 1 kg/kW in the next five years. Similarly the PV is about 7 kg/kW, or similar to the laser amplifiers. Interestingly, it is the radiator panels that are the most difficult to scale up, at about 25 kg/kW. This is an area that needs work, though in all simulations for mission masses the assumption is 25 kW (radiated) for the radiator panels.

ATK has to scale their existing 10 m diameter design to push the PV arrays to 30 m diameter which will yield about 225 kW per manufactured unit, or 450 kW per pair and still fit in an SLS PF1B 8.4 m diameter fairing. Fig. 9 and Fig. 10 show the scaling and deployment of the PV arrays to larger sizes for various launch vehicles. Even larger sizes into the megawatt range can be anticipated in the future.

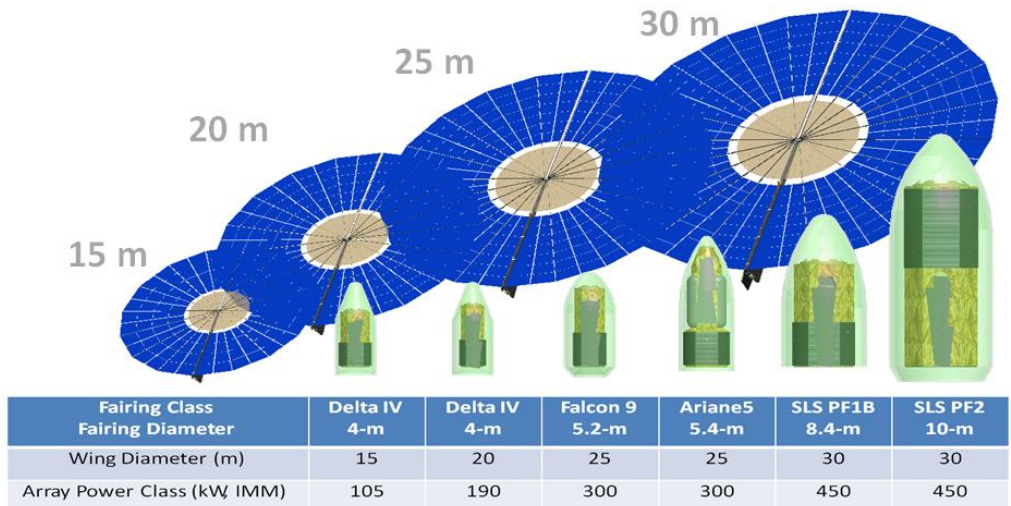

| Fairing Class Fairing Diameter | Delta IV 4-m | Delta IV 4-m | Falcon 9 5.2-m | Ariane5 5.4-m | SLS PF1B 8.4-m | SLS PF2 10-m |
|---|---|---|---|---|---|---|
| Wing Diameter (m) | 15 | 20 | 25 | 25 | 30 | 30 |
| Array Power Class (kW, IMM) | 105 | 190 | 300 | 300 | 450 | 450 |

**Figure 9.** Solar PV ATK Megflex arrays extended to 30 m diameter and 225 kW per panel for a total of 450 kW per pair. The 30 m diameter panel fits into the SLS fairing. Extension to the megawatt class could be accomplished with multiple units of these or possible extension to larger diameters.

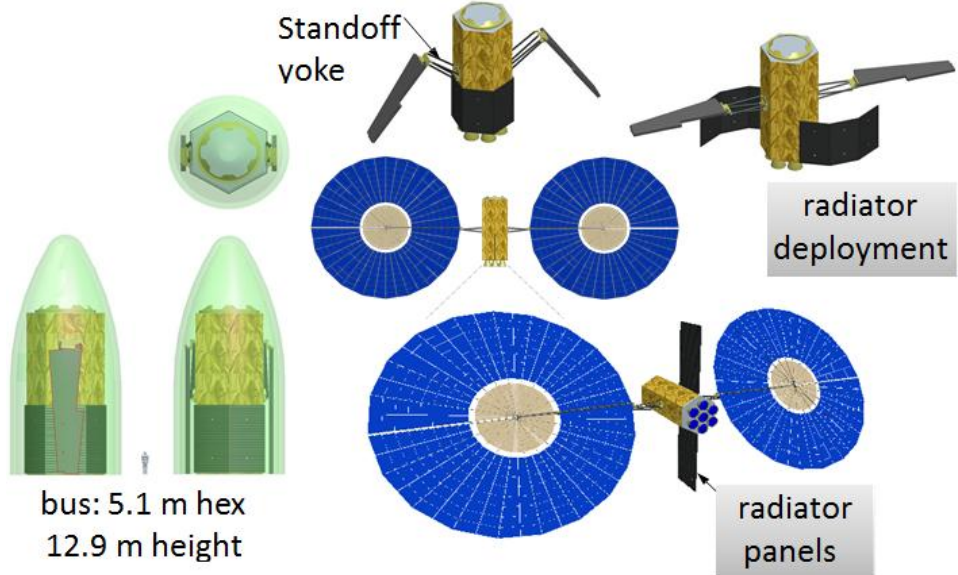

**Figure 10.** Deployment scenario for 30 m diameter 450 kW (pair) of ATK Megaflex panels from packing in an SLS fairing. Radiators are also shown as one possible option. Solar arrays are depicted larger than the emitter array, allowing for less efficient PV conversion.

## 5. ORBITAL DEFLECTION CAPABILITIES

### 5.1 How magnitude and duration of applied thrust influence miss distance

When an asteroid is exposed to the DE-STARLITE laser, the temperature [K] and flux [W/m$^2$] on the target asteroid must approach sufficiently high levels in order for significant ablation to occur, targeting a temperature on the order of 3000 K and a flux of >10$^7$ W/m$^2$. This causes direct evaporation of the asteroid at the spot of contact. Evaporation at the spot produces a vaporization plume thrust [N] that can be used to change the asteroid's orbit and effectively deflect asteroids from colliding with Earth. A miss distance of at least two Earth radii (12742 km) is required to eliminate the threat of collision. The orbital deflection depends on the duration, magnitude, and direction of the applied thrust.

Previous results describe analytical and semi-analytical treatments of orbital deflection. Colombo *et al.* (2009) use a semi-analytical approach to describe the motion of an asteroid subject to a low-thrust action with a thrust magnitude inversely proportional to the square of distance from the Sun. A simple low-thrust formula that shows the dependency of miss distance on $t^2$ is given in Scheeres and Schweickart (2004). Several previous papers explore optimal strategies for deflection Earth-approaching asteroids, including Conway (2001), Carusi *et al.* (2002), and Vasile and Colombo (2008). Colombo *et al.* (2009) provides a detailed derivation of the analytical formulae for orbit deflection as well as a comparison with full numerical simulations for different types of orbit. Zuiani *et al.* (2012) use first-order perturbation solutions of the accelerated motion to calculate an accurate deflection in the case of laser ablation. Bombardelli *et al.* (2011) develop asymptotic solutions for the deflection of asteroids with low thrust propulsion, including an analytical solution of the miss distance on the b-plane.

In this paper, we utilize a three-body simulation (accounting for the gravitational effects of the Earth, the sun, and the target asteroid) to analyze how the applied thrust and the laser-active time impact the miss distance (Zhang *et al.*, 2016; Zhang *et al.*, 2015-1; Zhang *et al.*, 2015-2). In order to determine the orbital deflection, Δx, of an asteroid that is being acted on over a period of time, t, an approximation that is commonly used in orbital mechanics was used as a comparison. The detailed numerical simulation is compared to the approximation of multiplying by 3 the naive distance achieved by accelerating and coasting a system that is not a bound gravitational system. Hence the orbital deflection is compared to:

$$\Delta x_{approx} = 3(0.5 \cdot a \cdot t_{active}^2 + a \cdot t_{active} \cdot t_{coast}) \quad (3)$$

where $a$ is the acceleration caused by the plume thrust, $t_{active}$ is the time the laser is active, and $t_{coast}$ is the coast time (typically zero). The reason this is done is because this approximation is often used for preliminary mission design. Note that compared to an impactor the deflection for the laser case (and other constant-force systems) scales quadratically with the time while the impactor case scales linearly. In reality, the actual orbital deflections are more complex.

The numerical simulations were performed in a rotating frame, where the thrust was pointed both along and against the velocity vector for comparison. Many dozens of orbital simulations were analyzed. The first data set shown compares the laser-active time to the miss distance for a given thrust acting on targets of varying diameter. This paper focuses on the 325 m diameter asteroid case, as this is approximately the size of Apophis—a well-known possible threat. Computations have also been done for 20 to 1000 m asteroids under many mission scenarios. The same code is used to analyze the IBD, gravity tractor and impactor (impulse) cases to which DE-STARLITE are compared. A sample of the results for the 325 m asteroid case is displayed in Fig. 11. It is evident that the factor-of-three approximation is indeed only an approximation and in some cases fails badly.

Further, for a given period of time, the required force is compared to the miss distance for targets of varying diameter. Simulations were run for targets of diameters between 20 m and 1000 m for varying mission parameters. Again, a sample of this data is displayed in Fig. 12 for a 325 m asteroid.

As is evident in Fig. 11 and Fig. 12 applying a thrust either parallel or antiparallel to the motion of the asteroid produces an equivalent deflection (but misses on opposite side of the Earth's orbit). Numerical analysis suggests that applying a thrust of 2.3 N (produced by ~100 kW electrical power @ 35% efficiency) over a period of 15 years will allow a 325 m asteroid to miss the Earth by two Earth radii. In contrast, if the laser were active for ten years, it would require approximately 5 N of force (produced by ~200 kW electrical power@ 35% efficiency) to deflect a 325 m asteroid by two Earth radii, and if it were only active for five years, it would require nearly 20 N of thrust (produced by ~870 kW electrical power@ 35% efficiency) to produce a comparable result.

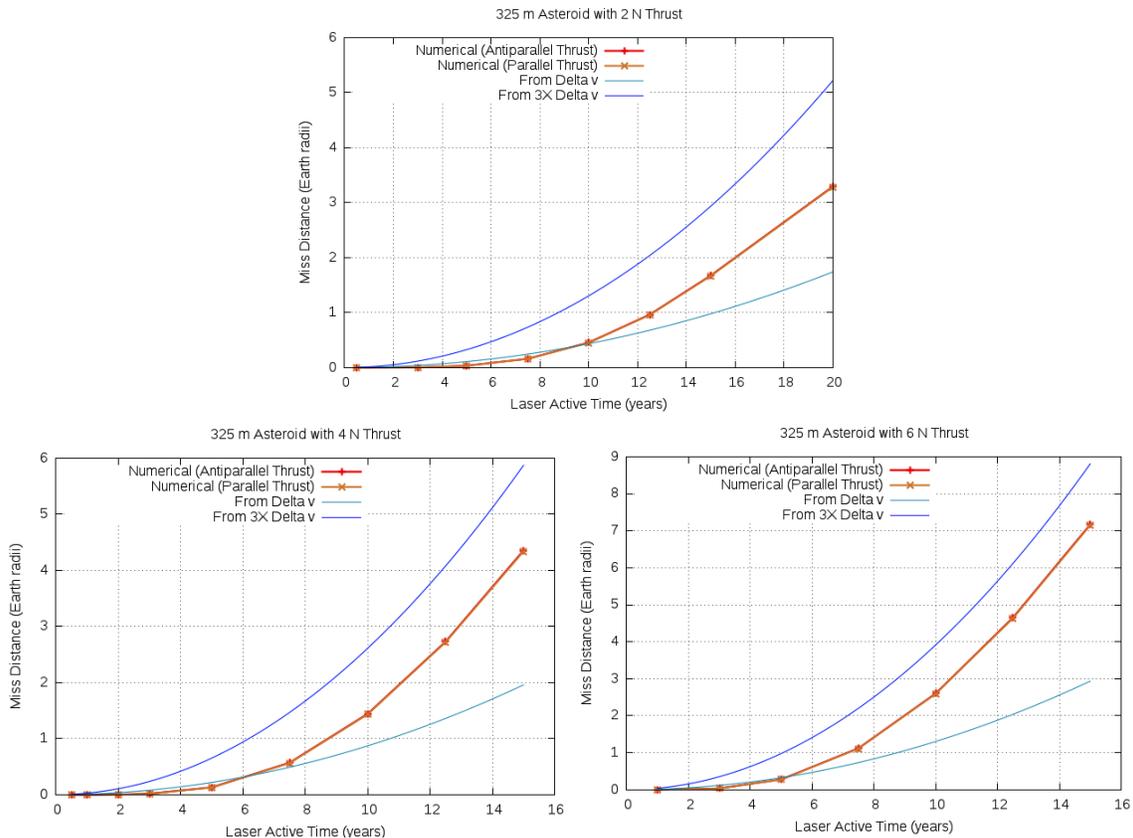

**Figure 11**. Plots of miss distance vs. laser-active time for an Apophis-sized asteroid (325 m) subject to a thrust of 2 N, 4 N, and 6 N, respectively. These thrusts are achievable with systems of approximately 20, 40 and 60 kW$_{optical}$ or about 50, 100 and 150 kW$_{electrical}$. Note that the analytic approximation and the detailed numerical modeling results approximately agree, but not in detail. This varies on a case-by-case basis, which is the point of being aware of the limitations of analytic approximations. See our other related figures in this paper.

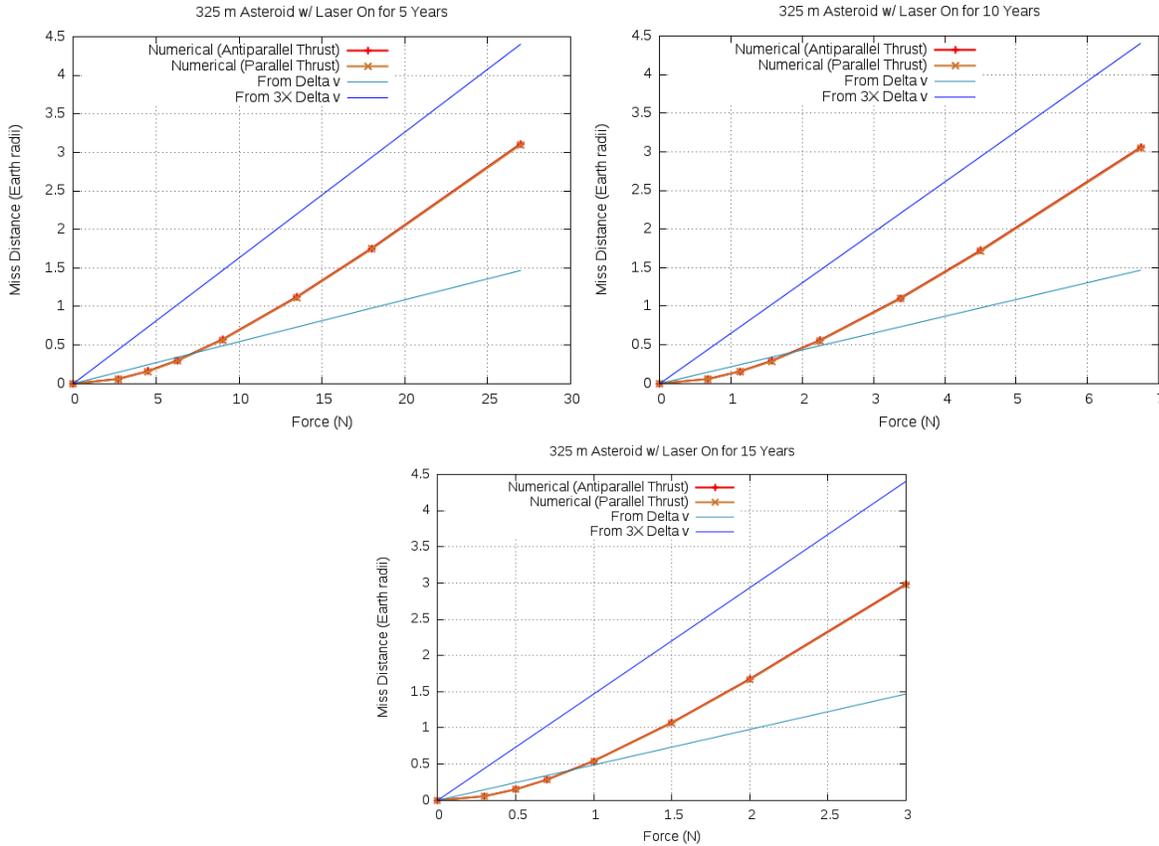

**Figure 12.** Plots of miss distance vs. thrust for an Apophis-sized asteroid (325 m) over a period of five years, ten years, and fifteen years, respectively. It is clear that operating over a longer period of time (longer warning time) greatly simplifies the system requirements in terms of thrust needed and power required. Note again the comparison between the analytic and numerical techniques.

**5.2 How optics diameter, distance from the target, and laser power effect the flux on the target**

For a laser phased array of a given power, a larger aperture enables the spacecraft to be further from the target while still producing sufficient ablation. The ratio of power to spot area must remain $>10^7$ W/m$^2$ in order for significant ablation to occur for high temperature compounds. Comets take much less flux due to their high volatility. With 35 kW of optical output, a laser phased array with a 4.5 m aperture can provide sufficient ablation from an approximate distance of 125 km, whereas a 35% efficient laser phased array with a 1 m aperture must be within 28 km of the target. Increased efficiency of the laser amplifiers will provide for even greater range. At 50% efficiency, with 50 kW of optical output, a 4.5 m laser array will have a range of approximately 150 km, while a 1 m array will have a range of roughly 33 km. Several cases are shown in Fig. 13.

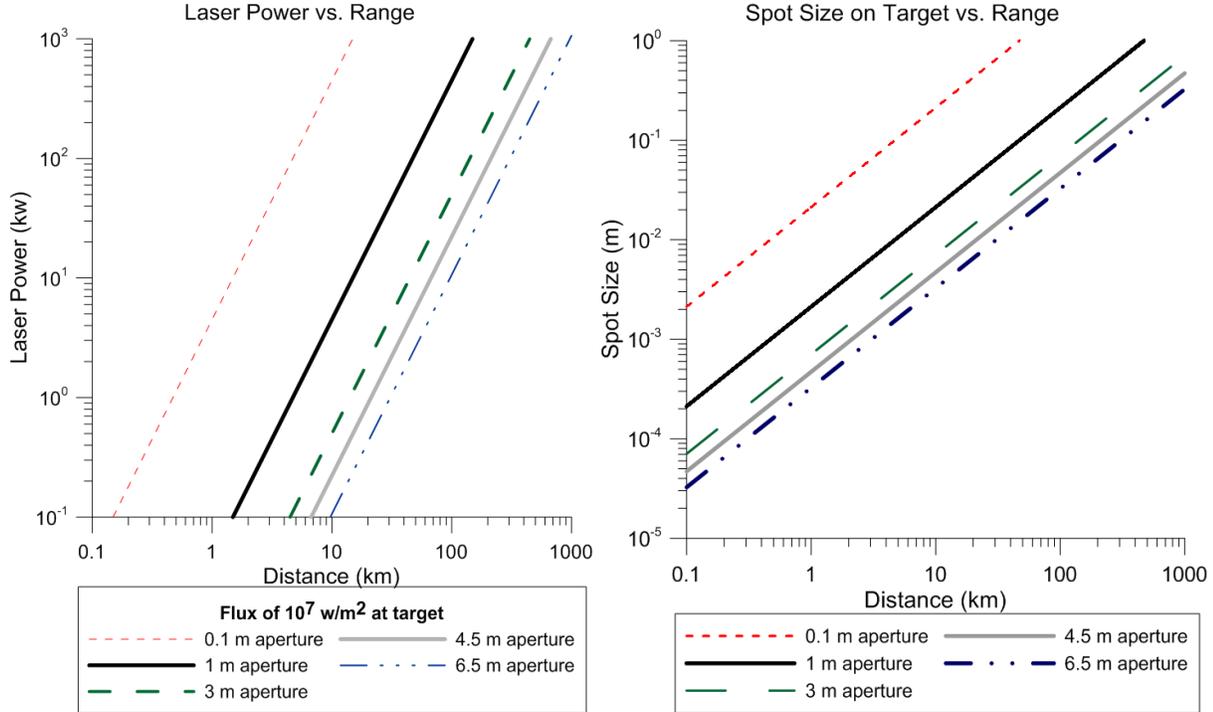

**Figure 13**. **LEFT**: Plot of laser power vs. range for various apertures. For the phased array, the relevant laser power is the sum of all the laser power, while for the close packed fiber array it is the power of EACH individual laser. **RIGHT**: Spot size vs. range for various apertures. The spot size on the target is approximately independent of whether a phased array or a close packed fiber array is used. It basically only depends on the aperture size and wavelength for a diffraction limited system.

For the close packed array alternative, the product of the optical diameter and the ratio of the spot sigma to the focal plane sigma provides an estimate for the target range. If a 1 m diameter optical system is implemented with a 2 mm spot sigma and 4 μm focal plane sigma, a range of 0.5 km can be achieved. Given the same parameters, a 4.5 m optical system will have a range of 2.3 km—much closer than with the phased array.

### 5.3 Comparison of efficiency of laser ablation versus ion beam deflection (IBD)

Ion beam deflection (IBD) and ion beam shepherding (IBS) is an alternative approach to achieve asteroid orbital deflection in which an ion beam / neutralized beam is used to push against the asteroid. In using this approach, the spacecraft must provide twice as much thrust as would otherwise be necessary to deflect the asteroid a desired distance. Half of the thrust is lost in station keeping in order to keep the spacecraft stable, as the spacecraft must push towards or away from the asteroid with an equal amount of thrust.

The mass required (Δm) for a desired impulse (Δp) is determined by the following succession of equations. The force $F$ is given by:

$$F = \frac{dm}{dt} V_{rel} \quad (4)$$

where $v_{rel}$ is the exhaust velocity. The desired impulse is

$$\Delta p = F \Delta t = \frac{dm}{dt} v_{rel} \Delta t = (\Delta m) v_{rel} \quad (5)$$

which can be solved to find the mass required for a desired impulse,

$$\Delta m = \frac{\Delta p}{V_{rel}} \quad (6)$$

The mass of propellant needed to produce an impulse of Δp on an asteroid using ion beam deflection is given by equation 7:

$$\Delta m = 2\frac{\Delta p}{V_{rel}} \qquad (7)$$

where the factor of two accounts for half of the thrust being used to stabilize the spacecraft. Using the IBD approach, the magnetically shielded Hall thruster is essentially reduced from 40 μN/W$_{elec}$ to 20 μN/W$_{elec}$ effective on the asteroid. Using Xenon as a propellant, the exhaust speed is effectively 30 km/s for a Hall effect thruster with an I$_{sp}$ of 3000 s. As a baseline example, to deflect an Apophis-sized asteroid (325 m) a miss distance of two Earth radii requires 2.3 N of thrust over a period of 15 years. According to equations 4-7, this action would require ~72300 kg of Xenon. An additional 5% of this mass is required to account for the tanks needed to hold the propellant, thus totaling 75900 kg. If gridded ion thrusters with an I$_{sp}$ of 6000 s are employed, which may be the case for a dedicated IBD mission, the mass of propellant required to produce a given thrust would be cut in half as the exhaust velocity is twice as great, though the power requirements roughly double in this case. The ion thrusters need to be chosen to have a small enough divergence angle and be close enough to the asteroid so most of the ions hit it.  While a significant amount of propellant is required to deflect an asteroid using the IBD method, no extra propellant is necessary after rendezvous with the asteroid (which uses a small amount of propellant from LEO) for the laser ablation case. A significant benefit of using laser ablation is that the asteroid is propelled by the ejection of its own material and thus the asteroid is itself the fuel. The mass of the laser array and the increased mass of the radiator (because the ion engines have a greater efficiency than the laser amplifiers) required for the laser ablation approach are of a far lesser magnitude than the additional mass of fuel needed for the IBD approach. Laser ablation is therefore proposed to be a more mass efficient mechanism by which to deflect an asteroid.

A laser ablation system such as the proposed DE-STARLITE system is much lower in mass than an equivalent IBD system. The following data presents preliminary mass estimates that should be treated as such. With roughly 7 tons to LEO, a DE-STARLITE spacecraft could accommodate 100 kW of electrical input, which corresponds to 50 kW of optical output (assuming 50% laser amplifier efficiency). The resulting effective thrust on the target is 2.5 N, given an effective thrust per electrical watt of 25 μN/W$_{elec}$. This assumes 50% laser efficiency, 70% of the optical power in the central spot (encircled energy) and 80 μN/W$_{optical}$ for the laser-asteroid coupling. Our simulations predict up to 5 times this amount; however, a conservative lower value is assumed here. Applying this amount of thrust over a period of 15 years would result in the orbital deflection of a 325 m asteroid by over two Earth radii. To produce the same result using IBD requires ~79900 kg (75900 kg for the propellant and tanks, as described above, as well as approximately 4000 kg for the dry mass of the spacecraft) if using magnetically shielded Hall effect thrusters, or ~42000 kg (approximately 38000 kg for propellant and tanks in addition to the dry mass of the spacecraft) if using gridded ion thrusters with an I$_{sp}$ of 6000 s. Note the higher I$_{SP}$ (6000 s) ion engine requires about twice the power of the lower I$_{SP}$ (3000 s) ARM engine for the same thrust. A trade study needs to be done to optimize this.

In comparing the systems, the extra ion engine fuel needed to deflect an asteroid could be instead used to massively increase the PV arrays and thus provide even more power to the laser system. Given 14 tons to LEO, a spacecraft utilizing IBD with Hall effect thrusters is estimated to be capable of outfitting 40 kW of electrical input, whereas a spacecraft utilizing laser ablation would be capable of supporting approximately 380 kW of electrical power by redistributing the mass. This corresponds to 190 kW of optical output for laser amplifiers operating at 50% efficiency, or a thrust on the target of  approximately 9.5 N—enough thrust to deflect a 250 m asteroid by a miss distance of two Earth radii over a period of five years. Thus with the same mass as a spacecraft that would be used to capture a 5-10 m asteroid, a system using laser ablation could protect the Earth from catastrophic devastation.

For a given power, the mass of the spacecraft utilizing laser ablation is approximately independent of the diameter of the target asteroid. Though the laser must be active for more time to deflect an asteroid of larger diameter, it does not require more mass to do so. In contrast, the mass of a spacecraft utilizing IBD increases as the cube of the asteroid diameter in order to accommodate more propellant to provide sufficient integrated thrust. The deflection time in both IBD and laser ablation (as well as most other approaches) increases with the cube of the asteroid diameter. Increasing the power output of the system will decrease the required warning time for a target of a given diameter because it will lessen the required laser-active time.  Several warning-time scenarios are depicted in Fig. 14. The DE-STARLITE mission calls for a baseline of 100 kW electrical, though this could be increased while staying within the mission parameters in order to decrease the required warning time. As shown above there is already a path to 30 m diameter class PV arrays that would yield about 450 kW of electrical power per pair. This path is consistent with the launch capabilities of the launch vehicles under consideration for the DE-STARLITE mission. Fig. 15 shows the estimate mission mass (at LEO) for various power scenarios while Fig. 16 shows the required laser on time vs. asteroid diameter. As is readily seen directed energy is extremely effective for even large targets with modest exposure times.

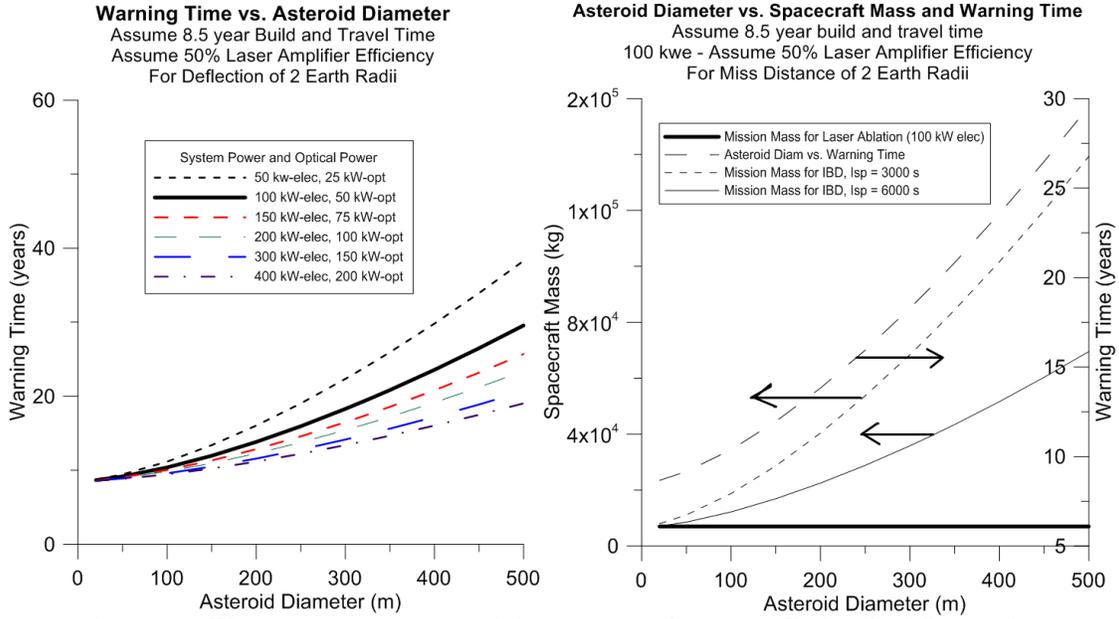

**Figure 14. LEFT**: Warning time versus asteroid diameter to produce a two Earth radii deflection for various system and optical powers assuming a laser amplifier efficiency of 50%, an 8.5 year build and travel time and an asteroid density of 2 g/cc. **RIGHT**: Asteroid diameter vs. spacecraft mass (left axis) for the IBD case (utilizing both magnetically shielded Hall effect thrusters with an $I_{sp}$ of 3000 s, and gridded ion thrusters with an $I_{sp}$ of 6000s) and for laser ablation, as well as asteroid diameter vs. the required warning time for a laser ablation system with 100 kW electrical power (right axis). For an equivalent warning time, the IBD case with an $I_{sp}$ of 3000 s requires ~125 kW electrical power, and the IBD case with an $I_{sp}$ of 6000 s requires ~250 kW electrical power. The same parameters (8.5 year build and travel time, 50% efficient laser amplifiers, 2g/cc and 2 Earth radii miss distance) are assumed. Note that the 8.5 year build and travel time is assumed for a spacecraft using ion engines with an $I_{sp}$ of 3000 s; the travel time may be decreased with ion engines of greater specific impulse and efficiency.

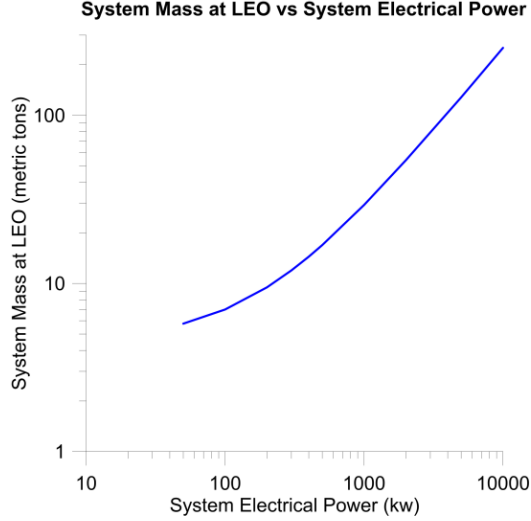
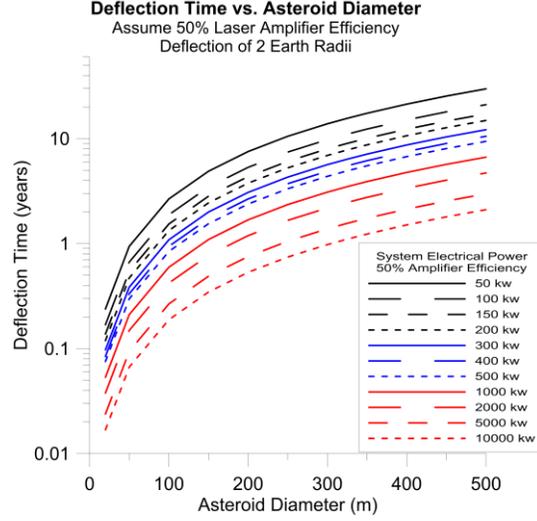

**Figure 15.** Spacecraft mass at LEO vs. electrical power available to system. Assumptions include the nominal ATK MegaFlex mass of about 7 kg/kW electrical, 5 kg/kW (optical power) laser amplifiers and 25 kg/kW (radiated) for the radiators and other nominal system bus parameters including ion engines and Xe fuel for LEO to asteroid. For reference an SLS Block 1 is spec'd at 70 metric tons to LEO and a Block 2 is 130 tons to LEO. Optical power from the laser is 1/2 of the electrical power for 50% efficiency laser amplifiers. Other electrical losses due to conversion and additional systems need to be included in a full analysis.

**Figure 16.** Laser deflection time needed to achieve 2 Earth radii miss distance vs. electrical power available assuming 50% amplifier efficiency and 80 $\mu N/W_{opt}$ coupling efficiency. Note that this is the laser on time not the warning time. The warning times in the figures above assume a build and travel time. The time shown here is the time the laser is actually on. The asteroid density is assumed to be 2000 kg/m$^3$.

### 5.4 Impactor Comparison

Here we discuss the case of using an impactor (ramming asteroid) vs. using a laser. As a common metric we use the launch mass as a common element for both cases–i.e., for the same launch mass, what can each system do? For a simplistic analysis the impactor delivers a large impulse or momentum transfer to deflect the target (integrated force - time in units of Ns). This momentum transfer imparts a change in the speed $\Delta V$ of the asteroid equals $\Delta p_{impactor}/M$ where $M$ is the mass of the asteroid. $\Delta p_{impactor}$ is the impulse delivered at a time $\tau$ before (if un-deflected) impact.

$\Delta p_{impactor} = \beta \cdot mv$ where $m$ is the spacecraft mass, $v$ is the relative closing speed between the spacecraft and asteroid and $\beta$ is an enhancement factor due to asteroid mass ejection from the impact. The enhancement factor is a much debated term that is a complex function of asteroid material properties at the impact site, geometry of impact, speed of impact. In general one assumes pure inelastic collision ($\beta =1$) to be conservative as any *a priori* $\beta$ is generally not going to be known for a given target. We will assume later that $\beta =1$ but it is important to keep this enhancement possibility in mind to be fair. The change of speed is thus:

$$\Delta V = \beta \cdot mv/M = \beta \cdot v \,(m/M) \quad (8)$$

The deflection distance at the Earth is approximately:

$$\Delta x_{impactor} = 3\, \Delta V \cdot \tau_{impact} = 3 \cdot \beta \cdot v \cdot \tau_{impact}\,(m/M) \quad (9)$$

where the factor of 3 is the same approximation used from orbital dynamics but as we have shown in several of our papers it is not always a good approximation (Zhang *et al*., 2015-1, Zhang *et al*., 2015-2) We use it here for analytic purposes for simplicity and because it is often used in mission planetary defense planning exercises. We also note that for impactors there can be an additional effect do the mass ejection upon impact. This depends on the asteroid materials, and the specifics of the impact.

Note that the miss distance $\Delta x_{impactor}$ is linearly proportional to the spacecraft or impactor mass ($m$), the closing speed ($v$) and time to impact $\tau$ and inversely proportional to the asteroid mass $M$. Note that the asteroid mass M is proportional to the cube of the asteroid diameter $D$. The momentum change (impulse delivered) is largely independent of the asteroid mass and only depends on the spacecraft mass ($m$) and the closing speed ($v$). For a homogeneous asteroid of density $\rho$ then miss distance is:

$$\Delta x_{impactor} = 3\ \Delta V \cdot \tau_{impact} = 18 \cdot m \cdot v \cdot \tau / (\pi \rho D^3) \qquad (10)$$

Since the asteroid is moving rapidly with typical speeds of 5-40 km/s we can simplify this to assume the spacecraft is simply in the way of the asteroid (inelastic billiard ball) and thus the speed of the spacecraft relation to the earth is of lesser importance. This of course depends on the specifics of the asteroid orbit (closing from the front vs. the back of the asteroid orbit). Essentially then it is the mass of the spacecraft that is critical to maximize. Once the space craft is launched to LEO it is assumed that ion engines will be used to allow a larger fraction of the launch mass to survive until impact to maximize the impulse. Since the miss distance is proportional to the inverse cube of the asteroid diameter, and the spacecraft mass is limited by the launcher capability, the only free parameter is the time to impact $\tau$. Thus the miss distance is:

$$\Delta x_{impactor} = 3 \cdot \Delta V \cdot \tau_{impact} = 3 \cdot \tau_{impact} \cdot \Delta p/M = 18 \cdot m \cdot v \cdot \tau_{impact} /(\pi \rho D^3) \qquad (11)$$

In other words, the miss distance is proportional to:

$$\Delta x_{impactor} \sim m \cdot v \cdot \tau \cdot D^{-3} \qquad (12)$$

For the case of directed energy the equivalent miss distance (using the same factor of 3 approximation for the effects of orbital mechanics) is (Chesley and Chodas, 2002):

$$\Delta x_{laser} = 3 \cdot 1/2 \cdot a \cdot \tau^2_{laser} = 3/2\ (a \cdot \tau)\ \tau_{laser} = 3/2\ \Delta V \cdot \tau_{laser} = 3/2\ (F/M)\ \tau^2_{laser}$$
$$= 3/2\ F \cdot \tau^2_{laser}/M = 1/2 \cdot 3 \cdot \Delta p_{laser} \cdot \tau_{laser}/M = 9\ \alpha\ P\ \varepsilon \cdot \tau^2_{laser} /(\pi \rho D^3) \qquad (13)$$

where:
- $a$ = acceleration imparted due to the laser plume thrust
- $F$ = laser plume thrust = $\alpha\ P\ \varepsilon$
- $P$ = laser power
- $\alpha$ = laser plume thrust coupling coefficient – thrust per optical watt
- $\varepsilon$ = beam efficiency factor – fraction of beam that is in central spot
- $\rho$ = asteroid density
- $D$ = asteroid diameter
- $M$ = asteroid mass = $\pi \rho D^3/6$
- $\tau_{laser}$ = laser ablation time (laser on time) – assumed to be on the entire time before impact and after rendezvous
- $\Delta p_{laser} = F\ \tau_{laser} = \alpha\ P\ \varepsilon\ \tau_{laser}$

Note that the laser deflection $\Delta x_{laser}$ is proportional to $\tau^2_{laser}$ while the impact deflection is proportional to $\tau_{impact}$. This is important as the deflection grows quadratically with time for the laser and linearly with time for the impactor.

We assume the laser thrust is constant and the asteroid mass changes very little due to the mass loss from ablation and that the laser plume thrust is proportional to the laser power. See our other papers on the detailed modeling for this. For simplicity we assume $\alpha \sim 80\ \mu N/W_{optical}$ based on our conservative laboratory measurements (Brashears *et al.*, 2015). These estimates are consistent with other published results (Gibbings *et al.*, 2013). Note that for the case of directed energy or any constant force (such as ion engines, gravity tractors, etc.) the miss distance:

$$\Delta x_{laser} = 1/2 \cdot 3 \cdot \tau_{laser} \cdot \Delta p_{laser}/M \qquad (14)$$

Where while for the impulse delivery (effectively instantaneously at a time $\tau_{impact}$ before impact) for the same overall delta momentum delivered to the asteroid is:

$$\Delta x_{impactor} = 3 \cdot \tau_{impact} \cdot \Delta p_{impactor}/M\ ,\ \text{or:}\ \Delta x_{laser} = 1/2\ \Delta x_{impactor} \qquad (15)$$

for the same $\Delta p$ and $\tau$. Again this is for the simplistic assumption of the factor of 3 to approximate the orbital mechanics effects. The question now becomes, "For a given launch mass which is more effective – impactor or laser?"

If we set the miss distance to be the same $\Delta x_{impactor} = \Delta x_{laser}$, then we can compare the laser-on time to impact-time, both before nominal Earth impact. We have

$$\Delta x_{laser} = 1/2 \cdot 3 \cdot \tau_{laser} \cdot \Delta p_{laser}/M = \Delta x_{impactor} = 3 \cdot \tau_{impact} \cdot \Delta p_{impactor}/M \qquad (16)$$

which gives:

$$\tau_{impact} / \tau_{laser} = 1/2\ \Delta p_{laser}/\Delta p_{impactor} = 1/2\ \alpha\ P\ \varepsilon\ \tau_{laser}/ \beta \cdot mv \qquad (17)$$

or

$$\tau_{impact} = 1/2\ \alpha\ P\ \varepsilon\ \tau^2_{laser}/ \beta \cdot mv \qquad (18)$$

Note that the ratios of times $\tau_{impact} / \tau_{laser}$ grows linearly with $\tau_{laser}$ so that the time ratio depends on the specifics of the case and not just on the fixed system parameters $\alpha$, $P$, $\varepsilon$, $\beta$, $m$, $v$. The real situation is far more complex than the 3 x delta approximations many times and depends on the specifics of the asteroid orbit and mission parameter as shown clearly in Fig. 17. We assume an SLS Block 1 launch of 70000 kg to LEO. For high Isp ion engines of 3000 s (Hall effect thrusters baselined for ARM) or 6000 s (gridded ion) a large fraction of the LEO mass will make it to the asteroid. We can show that for the **same mass limited launch the laser ablation system takes much less time to deflect the asteroid**. This is a critical point. Using the 3x delta analytic approximation we would conclude that for the same launch mass, but this time used for DE-STARLITE we would be able to launch a 1-2 MW laser system and for many scenarios this will be far more effective than an impactor of the same mass.

The details of the particular orbits are important but we can draw some basic conclusions. Assuming 60000 kg makes it out to the asteroid and with a closing speed of 10 km/s, the impactor impulse is $6 \times 10^8$ Ns. Fig. 17 shows that for this same 70000 kg SLS Block 1 to LEO, we could launch a 1 MW optical power laser delivering ~60 N of thrust on the asteroid for an assumed laser coupling coefficient α ~80 μN/W optical with an assumed beam efficiency in the central spot of 0.7. To get the same deflection in the same time to impact as the impactor, we need the laser system to deliver twice the momentum as the impactor since the impactor delivers the momentum change essentially instantaneously while the laser delivers it slowly over the entire time the laser is on. Hence, we need $1.2 \times 10^9$ N s. At 60 N of laser plume thrust this would require a time $\tau = 1.2 \times 10^9$ N s/60 N = $2 \times 10^7$ s **or about 7 months of laser exposure vs. using the same mass impactor which requires 10 years preemptive hit before Earth impact to obtain the same 2 Earth radii miss.** This time ratio depends on the specific of the asteroid orbit. Other differences for real systems are that typical impactor missions need more than one to make sure the impulse was delivered properly and that the asteroid orbital control with an impactor can be quite uncertain. For any real threat, multiple backups would be prudent.

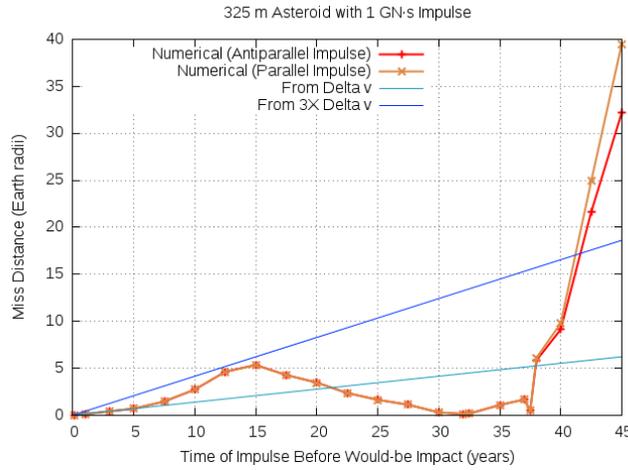

**Figure 17.** Miss distance vs. impulse delivery time before impact for 1 GN s impulse using an impactor on an Apophis class asteroid (325 m diameter) using both analytic approximations as well as detailed numerical simulations (Zhang *et al*., 2015-1; Zhang *et al*., 2015-2; Zhang *et al*., 2016). This impactor mass is somewhat larger than an SLS Block 1 can deliver. A miss distance of 2 Earth radii (typ. min acceptable) would require interdiction about 10 years before impact using this impactor. The seemingly unusual behavior of miss distance vs. time of impactor hit from the full simulation is due to resonance effects from the multiple orbits. It is clear the 3Δv approximation is not always accurate, and can be very misleading in some cases.

## 6. DIRECTED ENERGY AND ASTEROID ROTATION

### 6.1 Understanding Rotation Periods

All asteroids rotate, but generally quite slowly for larger one. A complete picture of rotation properties is not available, but from the limited data collected on the rotation of larger bodies and the break up speed it is estimated that asteroids in the 0.1-1 km class typically rotate no faster than once per several hours as seen in Fig. 18. Results of detailed observation indicate the rotation properties for more than 6000 significantly rotating asteroids and conclude that fast rotation is not an issue in general for larger asteroids (>150 m) as they are typically gravitational bound rubble piles (Walsh *et al*., 2012) and for these the maximum rotation is independent of diameter and only depends on density ρ, with an angular speed ω, and rotation period τ given by:

$$\omega = \sqrt{\frac{4}{3}\pi G \rho} = \frac{2\pi}{\tau}, \ \tau = \sqrt{\frac{3\pi}{G\rho}} \tag{19}$$

$$\tau \approx 1.19 \times 10^4 \rho[g/cc]^{-1/2} s \approx 3.3\rho^{-1/2}[hr], \text{ independent of diameter.} \tag{20}$$

Estimated densities are in the range of $\rho \sim 2$ [g/cc] yielding a minimum rotation period of about 2.3 hours. This is clearly seen in Fig. 18.

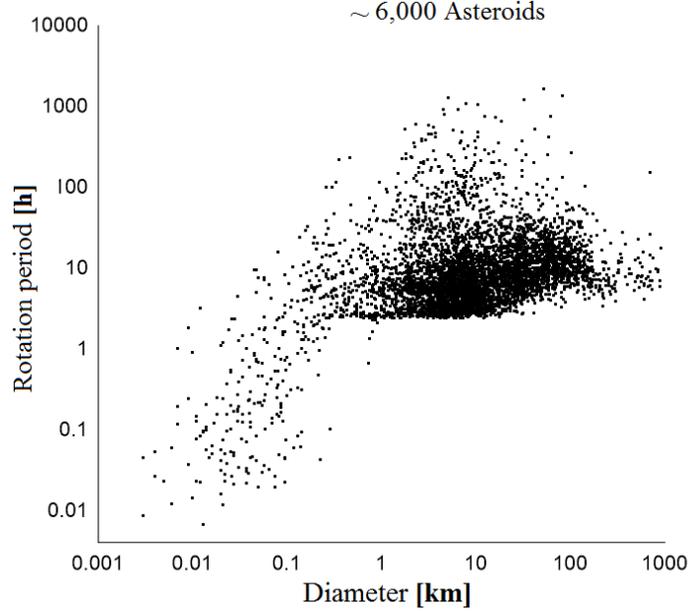

**Figure 18.** Measured rotation period of ~6000 asteroids. A distribution of measured asteroid rotation rates, notice the very sharp cutoff at just above 2 hours for larger diameter asteroids. Data from Minor Planet Center (Harris, 1998). The superfast rotators, those at the lower left with periods < 2.2 hours and D < 0.1 km are likely molecularly bound and form a distinct population.

The cutoff in rotation periods is observed to be remarkably sharp (see Fig. 18), and lies very close to 2 hours for asteroids of diameters greater than approximately 150 m, consistent with equation 20. Some smaller asteroids can rotate faster as they can have a tighter binding than purely gravitational (such as an iron meteorite) but these are relatively rare.

Even fast rotating asteroids can be dealt with since the mass ejection begins so quickly after the laser is turned on. As is seen in the transient thermal simulations below, the mass ejection and hence thrust begin within about 1 second. It is largely a flux issue so that for the same flux at any distance the mass ejection remains at this rate. This is assuming an asteroid consisting of solid $SiO_2$, which is extremely conservative. Loss is included to mimic the absorption qualities of asteroids, which are very absorptive having typical reflection coefficients around 5-10%. Thus, a rotating asteroid with this rate (1 hour) poses little problem. More interesting perhaps would be an attempt to spin up (or down) an asteroid depending on beam placement as discussed below. This is discussed in detail on one of our recent papers (Griswold *et al.*, 2015) along with laboratory measurements we made that show how effectively we can de-spin asteroids.

## 7. THERMAL ANALYSIS AND CURRENT MODELS
### 7.1 Comparison of Thermal Models
The thrust produced by DE-STARLITE on an asteroid is calculated using three different modeling approaches, of increasing complexity and realism. Results from the three analyses are compared, which all yield consistent answers. The basic equations are derived from energy conservation:

$$\text{Power in (laser)} = \text{Power out (radiation + mass ejection)} + \frac{dU}{dt}$$

Where U= Asteroid internal energy and $\frac{dU}{dt}$ is effectively from conduction.

In the steady state $\frac{dU}{dt} = 0$

$$P_{in} = P_{out} + \frac{dU}{dt}, \text{ with } U = \int \rho c_v dv \tag{21}$$

Where $c_v$ = specific heat [J/kg-K],
$F_L$ = Laser flux [W/m$^2$] - *(in)*, $F_{cond}$ = Thermal conduction [W/m$^2$] - *(in)*, and
$F_{rad}$ = Radiation flux [W/m$^2$] - *(out)*, $F_{ejecta}$ = Ejecta flux [W/m$^2$] - *(out)*.
Assuming $P_{in} = P_{rad} + P_{Ejecta} + P_{cond}$, then:

$$\oiint (\vec{F}_L - \vec{F}_{rad} - \vec{F}_{Ejecta} - \vec{F}_{cond}) \cdot \hat{n}\, dA = 0 \tag{22}$$

Or

$$\int \nabla \cdot (\vec{F}_L - \vec{F}_{rad} - \vec{F}_{Ejecta} - \vec{F}_{cond}) dV = 0 \tag{23}$$

Locally:

$$\vec{F}_L = \vec{F}_{rad} + \vec{F}_{Ejecta} + \vec{F}_{cond} \tag{24}$$

$$\vec{F}_{rad} = \sigma T^4 \hat{n} \tag{25}$$

$$\vec{F}_{Ejecta} = \Gamma e H_v \hat{n} = M^{1/2}(2\pi RT)^{-1/2} \alpha_e 10^{[A-B/(T+C)]} H_v \hat{n} \tag{26}$$

$$\begin{aligned} \left|\vec{F}_{cond}\right| &= K\nabla T, \\ \left|\vec{F}_{rad}\right| &= \sigma T^4, \text{ and } \left|\vec{F}_{Ejecta}\right| = \Gamma e \cdot H_v \end{aligned} \tag{27}$$

Where K is the thermal conductivity (which can be position and temperature dependent) and $\Gamma e$ is the mass ejection flux [kg/m$^2$-s], and H$_v$ is the heat of vaporization [J/kg]. The heat of fusion, H$_f$, is included for relevant cases. The heat of fusion is sometimes referred to the heat of sublimation as is sometimes the case for compounds in vacuum. H$_f$ is typically a small fraction of H$_v$. The mass ejection flux is shown in equation 28 which uses vapor pressure.

$$\Gamma e = \frac{M\alpha_e(P_v - P_h)}{\sqrt{2\pi MRT}} = M^{1/2}(2\pi RT)^{-1/2}\alpha_e(P_v - P_h) \tag{28}$$

Where:
- M = Molar mass $[kg/mol]$
- P$_v$ = Vapor pressure $[Pa]$
- P$_h$ = Ambient vapor pressure = 0 (in vacuum)
- $\alpha_e$ = coef. of evaporation

The models vapor pressure for each element and compound is determined using a semi analytic form known as Antoine coefficients A, B and C in equation 29.

$$LOG(P_v) = A - B/(T+C) \tag{29}$$

Where A, B and C are unique per element and compound. Hence:

$$P_v = 10^{[A-B/(T+C)]} \text{ and } \left|\vec{F}_{Ejecta}\right| = M^{1/2}\frac{1}{\sqrt{2\pi RT}}\alpha_e 10^{[A-B/(T+C)]} H_v \tag{30}$$

A Gaussian profile is assumed for the laser as an approximation shown in equation 30 where the Gaussian laser power is P$_T$, and r is the distance from the spot center.

$$\left|\vec{F}_L\right| = \frac{P_T}{2\pi\sigma^2} e^{-r^2/2\sigma^2} \tag{31}$$

In the approximation where the spot is small compared to the asteroid, the equation becomes:

$$\vec{F}_L = \frac{-P_T}{2\pi\sigma^2} e^{-r^2/2\sigma^2} \hat{n} \tag{32}$$

In the dynamic case, it is possible to solve for transient heat flow by:

$$\nabla \cdot (K\nabla T) + \frac{d}{dT}(\rho c_v T) = 0 \tag{33}$$

$$K\nabla^2 T + \rho c_v \frac{dT}{dt} = 0 \tag{34}$$

In equation 34, it is assumed that K (thermal conductivity) is independent of position, $\rho$ and $c_v$ are time independent. In the full 3D time dependent solution, all of the above conditions are invoked and the equations are solved simultaneously using a 3D numeric solver (COMSOL in this case). In the 2D steady state solutions, the thermal conductivity is assumed to be small (this is shown in 3D simulations to be a valid assumption as well as from first principle calculations) and a combination of radiation and mass ejection (phase change) is used:

$$\left|\vec{F_L}\right| = \left|\vec{F_{rad}}\right| + \left|\vec{F_{Ejecta}}\right| = F_T \tag{35}$$

$$F_T = \sigma T^4 + M^{1/2}(2\pi RT)^{-1/2} 10^{[A-B/(T+C)]} H_v \tag{36}$$

Inversion is not analytically tractable, so numerical inversion is used to get $T(F_T)$, which gives $P_v(F_T)$, $\Gamma_e(F_T)$, *etc*. In this inversion, a function fit is found (to 10$^{th}$ order typically):

$$T = \sum_{n=1}^{N} a_n (\log F_T)^n \tag{37}$$

A Gaussian approximation to the laser profile is used (this is not critical) to get $T(r)$, $P_v(r)$, $\Gamma_e(r)$ where r is the distance from the center of the spot.

Since radiation goes as the 4$^{th}$ power of T, while the mass ejection from evaporation goes roughly exponentially in T, at low flux levels the outward flow is completely dominated by radiation (the asteroid is heated slightly and it radiates). As the spot flux level increases (spot size shrinks or power increases or both) evaporation becomes increasingly dominant and eventually at about T ~2000-3000 K or fluxes of $10^6$ - $10^7$ W/m$^2$ mass ejection by evaporation becomes the dominant outward power flow and (just as water boiling on a stove) the temperature stabilizes and increasing flux only increases the rate of mass ejection with only very small increases in temperature.

The three methods:
- **1D Energetics alone.** Use heat of vaporization and set spot flux to correspond T ~6000 K *if* the system were completely radiation dominated. No radiation or conduction included, only vaporization.
- **2D Analytic** - Model elements and compound vapor pressure vs. T. Includes radiation emission. Ignore thermal conduction.
- **3D Numeric** - Full 3D FEA including phase change, vapor pressure, mass ejection, radiation and thermal conduction.

### 7.2 1D Energetics Alone

The heat of vaporization of a compound is the energy (*per mole or per kg*) to remove it from the bulk. Removal energy is related to an effective speed and an effective temperature, which are related to but somewhat different than the physical speed of ejection and the physical temperature of vaporization. To be more precise, the term evaporation refers to molecules or atoms escaping from the material (*for example water evaporating*), while boiling is the point at which the vapor pressure equals or exceeds the ambient pressure. At any non-zero temperature, there is a probability of escape from the surface: evaporation happens at all temperatures and hence vapor pressure is a quantitative measure of the rate of evaporation. The heat of vaporization is also temperature and pressure dependent to some extent. Table 2 gives thermal properties for various materials in asteroids. These materials have relatively high effective temperatures reflecting the fact that there is a probability distribution of energies and an increase in vapor pressure with respect to temperature (Lubin and Hughes, 2015).

| Material | $H_f$ [kJ/mol] | $H_v$ [kJ/mol] | M [g/mol] | $H_v$ [$10^6$ J/kg] | $C_v$ [J/kg-K] | $V_{eff}$ [km/s] | $T_{eff}$ [$10^4$ K] |
|---|---|---|---|---|---|---|---|
| **SiO$_2$** | **9.0** | **143** | **60.1** | **2.38** | **730** | **1.54** | **0.573** |
| Al$_2$O$_3$ | 14.2 | 293 | 102.0 | 2.87 | 930 | 1.69 | 1.15 |
| MgO | 77.4 | 331 | 40.3 | 8.21 | 1030 | 2.87 | 1.32 |
| ZnS | 38.0 | 320 | 97.5 | 2.46 | 472 | 1.57 | 1.28 |

**Table 2.** List of thermo-physical properties of common high temperature asteroid compounds. Here $H_f$ is the heat of fusion and $H_v$ is the heat of vaporization. $v_{eff} = \sqrt{H_v}$ [J/kg] and $T_{eff} = (M \cdot H_v)/3R$ where $R = k \cdot N_A \sim 8.31$

The thermal probability distribution has tail areas allowing for escape from the surface at lower temperatures than one would naively conclude from a mean analysis only. If power $P_T$ from the laser impinges on the asteroid in a small enough spot to heat to above the radiation dominated point (typically 2000-3000 K for rocky (monolithic) asteroids vs. 300-500 K for comets) it is possible to compute the evaporation flux (mass ejection rate) as: $M_e = P_T / H_v$. This is the maximum possible rate of mass ejection. It is possible to get quite close to this maximum if the system is designed properly.

### 7.3 2D Analytic

As mentioned above, this calculation assumes that the thermal conduction is small compared to radiation and mass ejection (a good assumption for most asteroids). Using the equations above and the numerical inversions it is possible to solve for the temperature distribution and thus the mass ejection and thrust on the asteroid among many other parameters. A summary is shown in Fig. 19 for $SiO_2$. The parameter σ (sigma) in the Gaussian beam profile is allowed to vary to show the effects of non-ideal beam formation as well as beam and pointing jitter. As can be seen the system is quite tolerant to errors in beam formation, focus, beam jitter and pointing errors even beyond 10σ as long as the power is high enough. The requirements on a low power system at equivalent distances are more severe. These relationships also show that it is possible to nearly achieve the theoretical maximum mass ejection rate. Also, note the thrust (N) per Watt is close to 0.001 N/W for the 1000 kW case. This is comparable to the Shuttle SRB in thrust per watt. This is not really surprising, considering that conventional propellants are approximately thermal in nature with temperatures close to the maximum sustainable in the combustion chamber and exhaust nozzle (*i.e.*, a few x$10^3$ K). More conservative numbers are assumed for system performance, typically 80 μN/W$_{optical}$ though calculations show the coupling to be between 100 and 500 μN/W$_{optical}$ depending on the asteroid material composition and the laser flux on target used (Riley *et al.*, 2014). More laboratory measurements are needed for various materials and flux levels. For now, a conservative value of 80 μN/W$_{optical}$ is assumed.

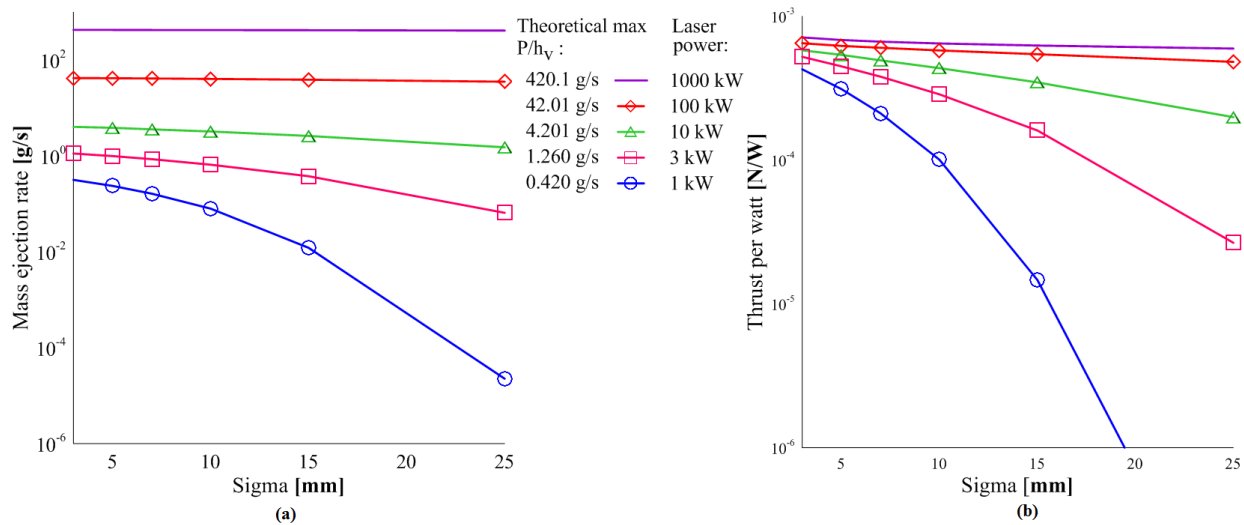

**Figure 19.** Using $SiO_2$ as the equivalent material. **(a)** Integrated mass ejection rates vs. sigma case for different powers between 1 kW and 1 MW. **(b)** Similarly, integrated thrust (N) per watt vs. sigma

### 7.4 3D Numeric Calculations and Simulations

Thousands of 3D model simulations have been run, and a few salient results are apparent. Calculations based on the simplest assumptions, namely energetics, and the conservation of spot flux, were validated. The more sophisticated tools are needed for further analysis and optimization of the system. For the case of dynamic targeting and rotating objects, time evolution has been added to the 3D solver. Some of this is motivated by the need to understand the time evolution of the mass ejection under dynamic situations. This is partially shown in Fig. 20 and Fig. 21, where the time evolution of the temperature at the center of the spot is shown. It is now possible to simulate full dynamics and apply this to the case of rotating asteroids. The same techniques can be applied to pointing jitter and laser machining (deliberate interior targeting) of the asteroid or other target.

The time evolution of the heated spot is shown in Fig. 21. Again, all cases refer to $SiO_2$ as the equivalent material for an asteroid. DE-STARLITE (as a stand-on system) is modeled here with a 1 m laser array, with a Gaussian beam and a total optical power of 1 MW, and spot diameter ~30 mm (σ ~5 mm). We use $SiO_2$ as a reference material; we have also run simulations for 92 elements and a number of compounds relevant to asteroid composition, including olivine family and other ultramafic minerals. All simulations produce essentially similar results for most of the appropriate compounds, within a factor of a few (Lubin and Hughes, 2015).

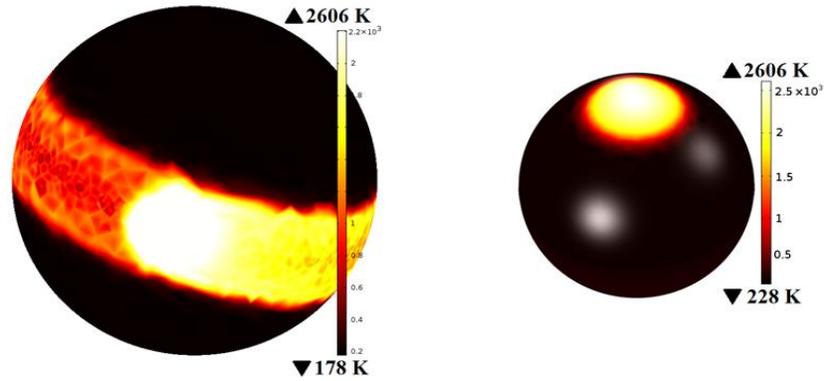

**Figure 20.** Rotating and stationary 3D plots for $SiO_2$: Using 1 hour rotation period for a 100 m diameter asteroid, yields equal surface temperature distribution as in the stationary steady state case. Temperatures rise to the point of being mass ejection limited, which is about 2600 K in the center of the spot. Solar illumination is modeled with an isotropic average of 350 W/m$^2$. The 1 hour rotation period is faster than the self-gravitating case and is shown as an extreme example of a large rotating asteroid that is not a rubble pile.

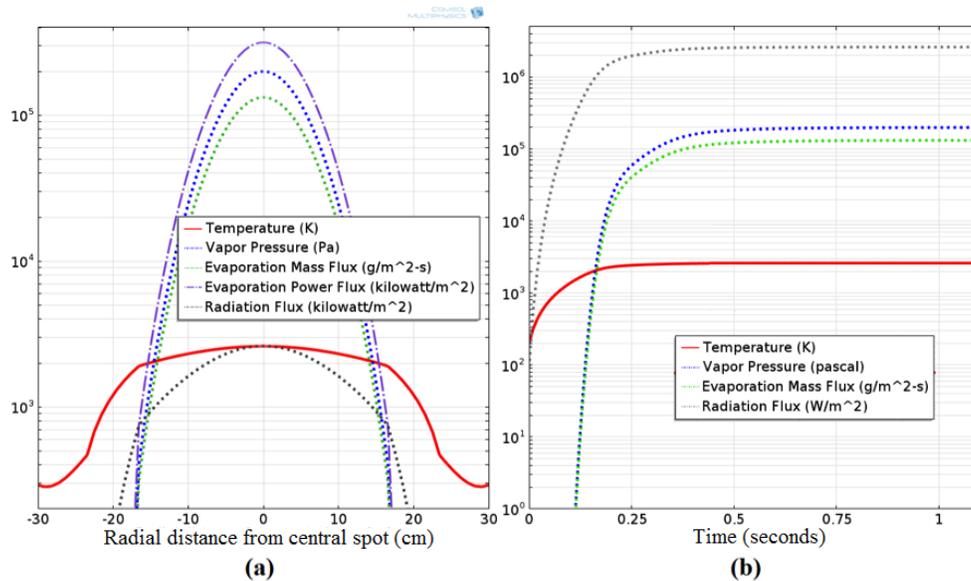

**Figure 21.** (a) Temperature, vapor pressure and mass loss distribution vs. distance from center (angle from beam axis). High frequency sub structure is due to numerical meshing. (b) Transient time solution (stationary) of temperature in the spot center (K) vs. Time (seconds) after the laser is turned on at $t = 0$. Initial temperature is 200 K. Mass ejection begins within 1 second. This case is for a DE-STARLITE with a 1 m optical aperture and 1 MW of optical power (this is a large DE-STARLITE) with a spot diameter ~30 mm (σ ~5 mm) on the target which is approximately 15 km away from the spacecraft. The same spacecraft could be over 100 km away from the target and still have about the same deflection.

## 7.5 Comparing Results Among Models

While the 3D simulations give time transient solutions and include full thermal conduction, they lack the numerical flexibility of the 2D solutions. Results of the temperature distributions for a Gaussian laser illumination are compared, and found to be very close in their predictions. This builds confidence that it is possible to do both 2D and 3D simulations with high fidelity. Fig. 22 shows comparisons of Gaussian beam illuminations; results are nearly identical in the critical central region.

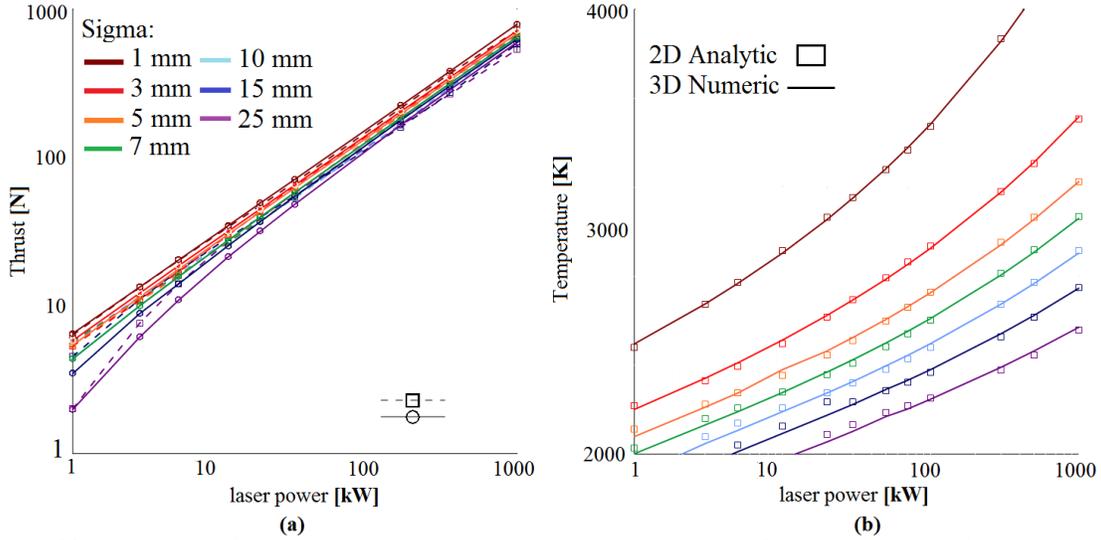

**Figure 22.** Comparison of 2- and 3-D models, hence numeric + analytic values for **(a)** integrated surface thrust (N) vs. total laser power for sigma between 1 and 25 mm. Note that the spot diameter (~6σ) for a DE-STARLITE kW class system is typically 3 to 75 mm. **(b)** Central spot temperatures. (In this case: DE-STARLINE – 1 m aperture)

The ultimate test will come when comparing model results with laboratory tests. As laboratory tests are refined, the results will feed back into the models for various materials.

## 7.6 Thermal Conduction

Unfortunately it is not possible to bring asteroids into the laboratory to study their thermal properties, so it is necessary to rely on astronomical observations, primarily in the infrared, combined with assumptions about their formation and likely structure, to deduce their properties. Several references (Mueller, 2007; Mueller *et al.*, 2007; Harris, 1998; Delbò *et al.*, 2007; Margot *et al.*, 2002), among many others, have done excellent work in this area and it is possible to use their results. One can derive the thermal properties by studying the time varying temperature as deduced from infrared observations. In this way the thermal inertia $\Gamma$ (J/m$^2$ K s$^{1/2}$) and thermal conductivity K [W/m K] are derived. The relationship between them is:

$$\Gamma = [\rho \, K \, C]^{1/2} \tag{38}$$

Where:
   $\rho$ = Density [kg/m$^3$]
   C = heat capacity [J/kg K]
Hence:

$$K = \Gamma^2 / (\rho \cdot C) \tag{39}$$

The data is shown in Fig. 23 best fit to published data (Delbò *et al.*, 2007), where D is the asteroid diameter [km] is:

$$\Gamma = d \cdot D^{-\zeta} \tag{40}$$

With d = 300 [km], $\xi$ = 0.4, and

$$K = 3e4 \cdot D^{-0.8} / (\rho \cdot C) \tag{41}$$

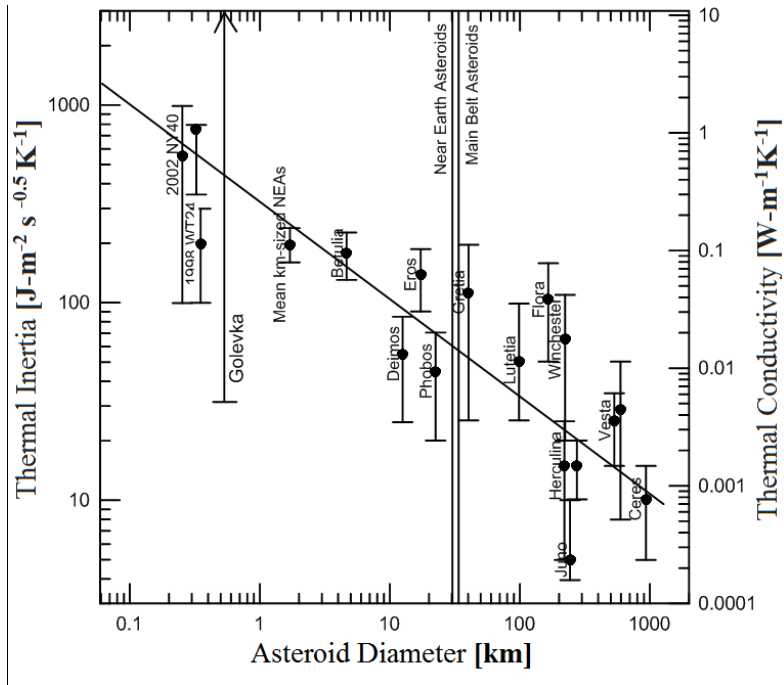

**Figure 23.** Thermal Inertia Γ - [J/m² K s^(1/2)] and Thermal Conductivity: K [W/m K]

The trend (*with some significant deviations*) is towards smaller asteroids having larger thermal conductivity and larger asteroids having smaller thermal conductivity as shown in Fig. 23. Some of this may be the point contacts from rubble-pile effect for larger asteroids. A similar trend between asteroid size and thermal inertia is also observed. It is the values that are of interest in the models. A relatively conservative case of K = 1 [W/m K] is assumed. To put this in perspective, some values for common materials are given in Table 3.

| Material | K [W/m K] | ρ [kg/m³] | C [J/kg K] | Γ [J/m² K s^(1/2)] |
|---|---|---|---|---|
| Nickel | 91 | 8850 | 448 | 1.9x10⁴ |
| Iron | 81 | 7860 | 452 | 1.7x10⁴ |
| Granite | 2.9 | 2750 | 890 | 2600 |
| Ice (solid) | 2.3 | 917 | 2000 | 2040 |
| **SiO₂ (solid)** | **1.04 (*at 200 °C*)** | **2200** | **1000** | **1510** |
| Water (liq 0°C) | 0.56 | 1000 | 4200 | 1500 |
| Snow (firm) | 0.46 | 560 | 2100 | 740 |
| Soil (sandy) | 0.27 | 1650 | 800 | 600 |
| Pumice | 0.15 | 800 | 900 (*varies significantly*) | 330 |
| Styrofoam | 0.03 | 50 | 1500 | 47 |
| Air | 0.026 | 1.2 | 1000 | 5.6 |
| Moon (regolith) | 0.0029 | 1400 | 640 | 51 |

**Table 3.** Common material thermal properties for comparison to the asteroid thermal properties in Fig. 24

Raising laser power from 10 kW to 20 kW resulted in slightly smaller range between minimum and maximum final temperatures with a relatively small effect on the final temperature between the two laser powers. This is to be expected since the effective vapor pressure and hence mass ejection rate and hence power into mass ejection is a strong function of the temperature. For these simulations, a relatively conservative case of K= 1 W/m K is assumed. For values of thermal conductivity between 0.01 and 250 W/m K, the evaporation mass flux and thrust change only slightly, shown in Fig. 24.

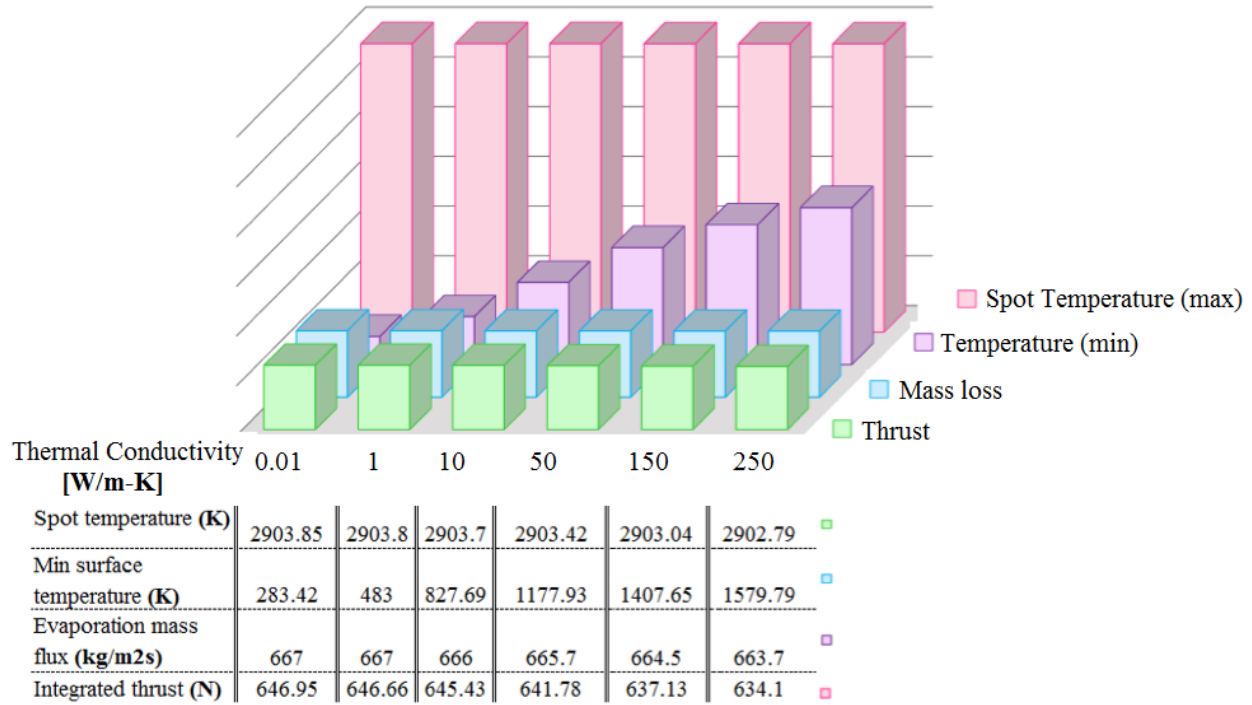

**Figure 24.** Extreme values inputs of thermal conductivity set to 0.01-250 W/m K for $SiO_2$ – Using 1 MW laser power, spot diameter is 60 mm, with sigma 10 mm, in this case for a 2 m diameter asteroid.

## 8. EFFECTS OF ROTATION ON SYSTEM REQUIREMENTS

From the simulations shown in Figure 20 it is clear that the mass ejection process begin rather quickly, typically within a second of laser initiation. The time scale for mass ejection is also dependent on thermal conductivity, density, heat capacity and heat of vaporization. It is possible to make an estimate of the effects of asteroid rotation by considering the effective motion of the laser spot in the worst case of the spot on the equator. This is shown in Fig. 25. The spot will then move relative to the asteroid surface at a surface speed determined by both the rotation period and the diameter of the asteroid. One simple way to think about the relative time scales is to compare to mass ejection time (after laser initiation) to the time to move the laser spot by about one spot size. If the spot moves a large amount (compared to the spot size) in the time it takes to begin mass ejection then the system will be seriously compromised in terms of effectiveness. The bottom line is that faster rotating asteroid need higher power levels and slower can use lower power. A possible solution to reduce the average power is to use the laser in a pulsed (higher peak power) mode to de-spin it and then run CW to deflect it to optimize the lowest possible average power needed. The pulsed high peak power mode allows for higher flux so spot smearing effects are not as important and allows the target to be spun down to near zero rate relative to the velocity vector. Once the asteroid is spinning slowly enough the CW laser mode can start for full deflection capability. The next section discusses using directed energy to de-spin the asteroid.

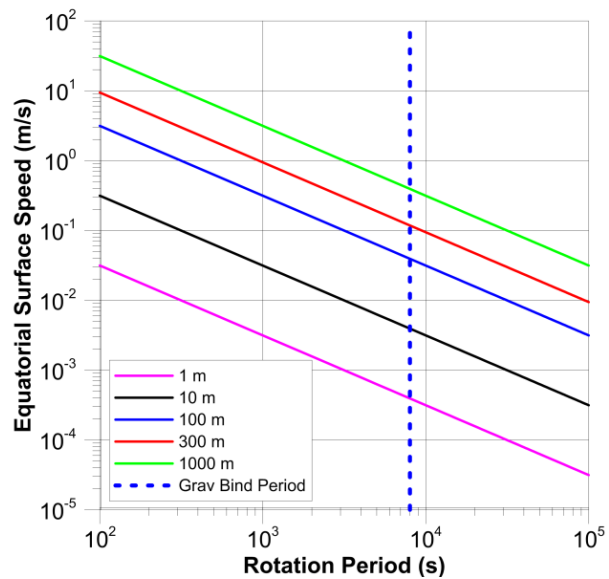
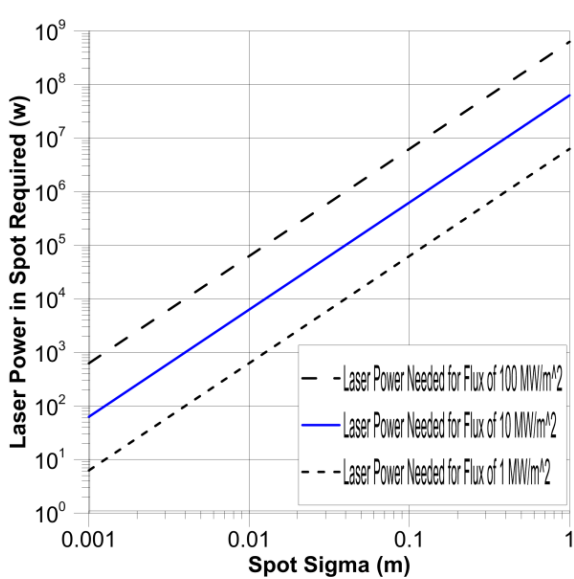

**Figure 25. Left:** Laser spot surface speed at the equator vs. rotation period vs. diameter of the asteroid. The 2.3 hour gravitational binding limit (rubble pile limit) is shown for reference for a density = 2 g/cc asteroid. **Right:** Laser power needed vs. the spot sigma for a Gaussian laser beam for three different flux level requirements. Typically 10 MW/m² is sufficient flux for most materials to be efficient at mass ejection. The effective spot diameter can be estimated as ~6 σ depending on how one views a Gaussian beam to spot diameter conversion. The way to think about this is you want the surface speed (Left figure) to be less than one spot diameter per second (since the time to mass ejection is typically around 1 s or less). Knowing the spot speed you can then determine the laser power needed to reach the flux required (Right figure). This then drives you to larger power levels for larger diameter asteroid for the same rotation period. As an example to effectively work with a 100 m diameter asteroid that is a rubble pile (2.3 hour rotation rate) you have about a 5 cm/s spot speed. This then requires a spot sigma around 1-5 cm which requires a power level of about 10-60 kW. For a 300 m diameter asteroid rotating at the rubble limit you have a 12 cm/s spot speed and need a spot sigma of 2-10 cm with a power level of 30-700 kW. Slower rotating asteroids need less power and faster ones need more. Running the laser array in an optional pulsed high power mode (short duty cycle so average power remains the same) can overcome this problem allowing the asteroid to be de-spun first and then fully deflected while running in CW mode. See Griswold *et al.* (2015) for details.

## 9. ASTEROID ROTATION MITIGATION

### 9.1 De-spinning a Rotating Asteroid

With laser ablation technology it is possible to change the spin of an asteroid. The small spot and fine control allow the ability to do precision manipulation on a target. This could be useful in de-spinning an asteroid for capture, landing or mining missions as examples. The time it takes to de-spin an asteroid depends on thrust (torque), initial angular velocity and asteroid diameter. Simple calculations allow calculating the torque necessary to de-spin a rotating spherical solid, assuming homogeneous composition and density. The torque can be varied by changing the power level, changing the spot size or moving the spot to different locations relative to the spin axis is shown below. To be able to spin down or spin up an asteroid is one of the unique abilities of a directed energy system. If 1000 hours (about 40 days) are allotted to spin down a 150 m diameter asteroid that has an initial period at the gravitation binding limit (about 2.3 hours) it requires about 20 N of thrust to do so. This would require a fairly large system with about 200 kW of optical power. One option to reduce the required power is to allot more time to spin down (this depends on the threat time to impact) - for example allotting a year to de-spin this asteroid would only require about 2 N of thrust or about 20 kW of optical power. The optimization requires specific details of the threat parameters but de-spinning an asteroid remains an interesting option for a directed energy system. It is possible to derive the time required to de-spin a rotating asteroid by modeling the system below. Assumptions are that the laser power and thus the flux and hence mass ejection is constant over the time used to de-spin and that the illuminated spot is at a constant location relative to the spin axis. The worst case of the spin axis perpendicular to the velocity axis is also assumed. In general the spacecraft will be aligned (or anti aligned) along the velocity axis. The mass loss during this time is assumed to be minimal compared to the total asteroid mass. For these assumptions the torque from the ejection plume is constant and thus the angular acceleration α is constant. In practice a real system will be more complex for many reasons as discussed below but this gives us a first order solution. Solutions are plotted in Fig. 26 and Fig. 27. Consider a rotating asteroid with the following parameters:

$P = 2000$ kg/m$^3$
$T$ = initial rotational period (s)
$\omega_0 = 2\pi/T$ initial rotation speed (rad s$^{-1}$)
$t$ = desired time to stop rotation (s)
$L$ = lever arm, $0 < L < R$ (m)
$R$ = Asteroid Radius (m)

The sum of torques on the asteroid is:

$$\sum \tau = I\alpha = I\frac{\partial \omega}{\partial t} \tag{42}$$

where:

$$|\tau| \equiv |F \times L| \leq F \cdot R \cdot \sin(90) = |I\alpha| \tag{43}$$

To de-spin the asteroid, the final rotational speed must be zero:

$$\omega_{final} = \omega_0 - \alpha t = 0 \tag{44}$$

hence:

$$\alpha = \frac{\omega_0}{t} \tag{45}$$

For a solid sphere:

$$I = \frac{2}{5}MR^2 \to M = \rho \cdot \frac{4}{3}\pi R^3 \to \to I = \frac{2}{5}\left(\rho \cdot \frac{4}{3}\pi R^3\right)R^2 \text{ or } \rho\frac{8\pi R^5}{15} \tag{46}$$

The torque $\tau(N - m)$ vs. T becomes:

$$\tau(T) = I\frac{\omega_0}{t} = \left(\rho\frac{8\pi R^5}{15}\right)\cdot\left(\frac{2\pi}{t \cdot T}\right) = \frac{\rho \cdot \frac{16}{15}\pi^2 R^5}{t \cdot T} \tag{47}$$

The required thrust F to spin down (stop rotation) for $L = R$ becomes:

$$F(T) = \frac{\tau(T)}{L} \approx \left(\frac{8\pi}{15}\rho R^5 \frac{\omega_0}{t}\right)\left(\frac{1}{R}\right) \approx \frac{\rho \cdot \frac{16}{15}\pi^2 R^4}{t \cdot T} \tag{48}$$

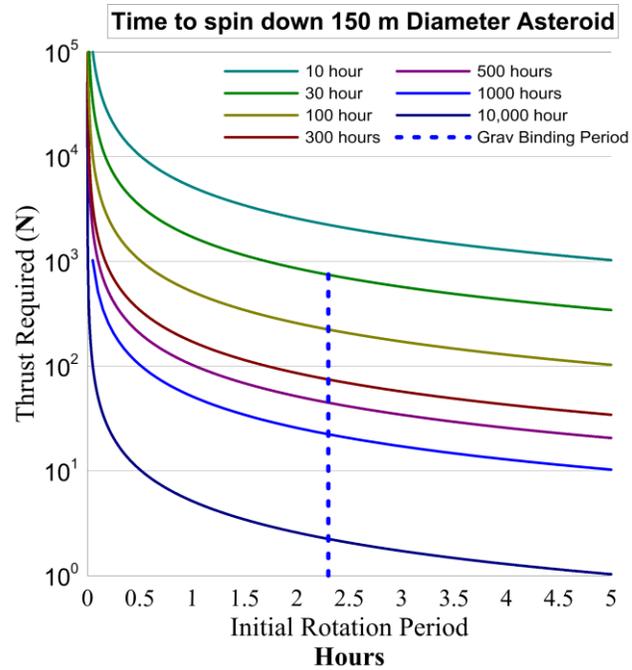
**Figure 26.** Thrust required as a function of rotation period (hours) to de-spin a 150 m diameter asteroid with a density of 2000 kg/m³. As an example to spin down a 150 m diameter asteroid, that is rotating at a period of about 5 hours, in 1000 hours of illumination (about 40 days) takes about 10 N of thrust.

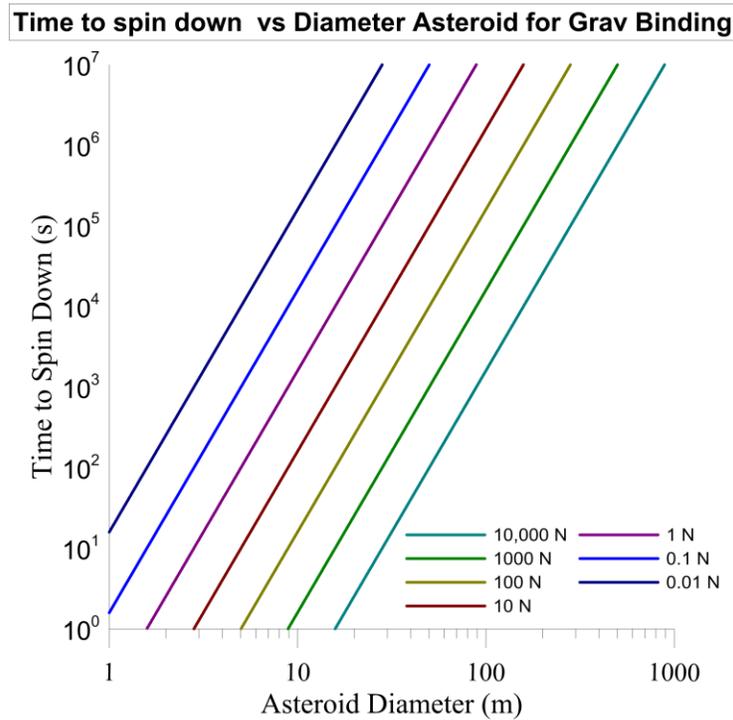
**Figure 27.** Time to spin down as a function of asteroid diameter and toque applied assuming the asteroid is spinning at the gravitational binding rate (~2.3 hours) with a density of 2000 kg/m³.

**9.2 Some Future Work in Rotational Studies**

- Additional 4D simulations with beam size and flux with varying rotation rates.
- Asteroids that are smaller than ~100 km in diameter are rarely close to spherical. It will be necessary to run simulations that are for non-spherical geometries.
- Run heterogeneous composition models with a rubble surface and regolith like coating.
- Shapes and binary systems (Margot *et al*., 2002): Asteroids like stars, comes in multiplicity (13% NEA's - Near Earth Asteroids/deflected from the main belt).
- Precession, Synchronic, and Chaotic motion
- Run full simulations including rotation with orbital dynamics.
- Simulations that combine pulsed and continuous (CW) modes to look at optimization.

## 10. CONCLUSIONS

    The DE-STARLITE system provides a feasible solution to asteroids and comets that pose a threat to Earth. By utilizing a directed energy approach with a high powered phase locked laser array to vaporize the target surface the thrust generated from the mass ejection plume is able to propel the asteroid threat away from the original collision trajectory towards Earth. DE-STARLITE is a very system at a modest cost. As outlined above, DE-STARLITE employs laser ablation technologies which use the asteroid as the propellant source for its own deflection, and thus is able to mitigate much larger targets than would be possible with other proposed technologies such as IBD, gravity tractors, and kinetic impactors. With the equivalent mass of an ARM Block 1 arrangement (14 tons to LEO - full SLS block 1 is 70 tons to LEO), designed to capture a 5-10 m diameter asteroid, DE-STARLITE can mitigate an asteroid larger than Apophis (325 m diameter), even without keyhole effects. Much smaller DE-STARLITE systems could be used for testing on targets that are likely to pass through keyholes. The same technology proposed for DE-STARLITE has significant long-range implications for space missions, as outlined in other DE-STAR papers. Among other benefits, the DE-STARLITE system utilizes rapidly developing technologies to perform a task previously thought to be mere science fiction and can easily be increased or decreased in scope given its scalable and modular nature. DE-STARLITE is capable of launching on an Atlas V 551, Falcon Heavy, SLS, Ariane V or Delta IV Heavy, among others. Many of the items needed for the DE-STARLITE system currently have high TRL; however, one critical issue currently being worked on is the radiation hardening of the lasers, though it appears achievable to raise this to a TRL 6 within 3-5 years. Laser lifetime also poses an issue, though this is likewise being worked on; a path forward for continuous operation looks quite feasible, with or without redundancy options for the lasers. Given that the laser amplifier mass is small and the system is designed to take multiple fibers in each configuration, redundant amplifiers can be easily implemented if needed. DE-STARLITE is a critical step towards achieving the long-term goal of implementing a standoff system capable of full planetary defense and many other tasks including spacecraft propulsion. DE-STARLITE represents a practicable technology that can be implemented within a much shorter time frame at a much lower cost. DE-STARLITE will help to establish the viability of many of the critical technologies for future use in larger systems.

    Since all asteroids rotate at varying rates, this will cause the average applied thrust to decrease and this must be taken into account in the system design. A lower limiting rotation period for gravitationally bound objects greater than 150 m is observed to be 2-3 hours consistent with being rubble piles. This effect needs to be taken into account for larger asteroids and for small fast rotators. Since the plume thrust begins within 1 second after the laser is initiated it is possible to compare the time scales of the laser spot motion to the mass ejection time scale to determine the effect of the rotation. In many cases rotation is not a fundamental concern but for those cases where it is, an option is to de-spin the asteroid, since this is an option with the proposed system. Running in a high peak power pulsed mode is one option available to mitigate rotation and allow de-spin. In summary, directed energy is an extremely promising option for true planetary defense. It is modular and scalable and allows for a very cost effective approach that has wide applications beyond planetary defense.

## ACKNOWLEDGEMENTS

We gratefully acknowledge funding from the NASA California Space Grant NASA NNX10AT93H in support of this research.